\documentclass{raa07}
\usepackage{graphicx,times}
\usepackage{natbib}
\usepackage{amssymb,amsmath}
\usepackage{booktabs}
\usepackage[figuresright]{rotating}

\bibpunct{(}{)}{;}{a}{}{,}

\usepackage[pagebackref=true]{hyperref}
\hypersetup{pdftitle = The title of my PDF, pdfauthor = My name, pdfsubject= The subject, pdfkeywords = keyword1 keyword2 keyword3}
\hypersetup{colorlinks = true, linkcolor = green, anchorcolor = red, citecolor = blue, filecolor = red, urlcolor = red}

\begin{document}

   \title{LAMOST Time-Domain Survey: First Results of four $K$2 plates}

   \setcounter{page}{1}

   \author{Song Wang\inst{1}, Haotong Zhang\inst{1}, Zhongrui Bai\inst{1}, Hailong Yuan\inst{1}, Maosheng Xiang\inst{2}, Bo Zhang\inst{3}, Wen Hou\inst{1}, Fang Zuo\inst{1}, Bing Du\inst{1}, Tanda Li\inst{4,5,3}, Fan Yang\inst{3,1,6}, Kaiming Cui\inst{1,6}, Yilun Wang\inst{1,6}, Jiao Li\inst{1}, Mikhail Kovalev \inst{7,2}, Chunqian Li\inst{1}, Hao Tian\inst{1}, Weikai Zong\inst{3}, Henggeng Han\inst{1,6}, Chao Liu\inst{8,6}, A-Li Luo\inst{1,6}, Jianrong Shi\inst{1,6}, Jian-Ning Fu\inst{3}, Shaolan Bi\inst{3}, Zhanwen Han \inst{7}, Jifeng Liu\inst{1,6,9}}

   \institute{ Key Laboratory of Optical Astronomy, National Astronomical Observatories,
Chinese Academy of Sciences, Beijing 100101, China; {\it  songw@bao.ac.cn}\\
\and
Max-Planck Institute for Astronomy, K{\"o}nigstuhl 17, D-69117 Heidelberg, Germany\\
\and
Department of Astronomy, Beijing Normal University, Beijing 100875, China\\
\and
School of Physics and Astronomy, University of Birmingham, Birmingham B15 2TT, UK\\
\and
Stellar Astrophysics Centre, Department of Physics and Astronomy, Aarhus University, Ny Munkegade 120, DK-8000 Aarhus C, Denmark\\
\and
College of Astronomy and Space Sciences, University of Chinese Academy of Sciences, Beijing 100049, China\\
\and
Yunnan Astronomical Observatory, China Academy of Sciences, Kunming, 650216, China\\
\and
Key Lab of Space Astronomy and Technology, National Astronomical Observatories, Chinese Academy of Sciences, Beijing 100101, China\\
\and
WHU-NAOC Joint Center for Astronomy, Wuhan University, Wuhan, Hubei 430072, China\\
\vs \no
   {\small Received 20XX Month Day; accepted 20XX Month Day}
}

\abstract{
From Oct. 2019 to Apr. 2020, LAMOST performs a time-domain spectroscopic survey of four $K$2 plates with both low- and med-resolution observations.
The low-resolution spectroscopic survey gains 282 exposures ($\approx$46.6 hours) over 25 nights, yielding a total of about 767,000 spectra, and the med-resolution survey takes 177 exposures ($\approx$49.1 hours) over 27 nights, collecting about 478,000 spectra.
More than 70\%/50\% of low-resolution/med-resolution spectra have signal-to-noise ratio higher than 10.
We determine stellar parameters (e.g., $T_{\rm eff}$, log$g$, [Fe/H]) and radial velocity (RV) with different methods, including LASP, DD-Payne, and SLAM.
In general, these parameter estimations from different methods show good agreement, and the
stellar parameter values are consistent with those of APOGEE.
We use the {\it Gaia} DR2 RV data to calculate a median RV zero point (RVZP) for each spectrograph exposure by exposure, and the RVZP-corrected RVs agree well with the APOGEE data.
The stellar evolutionary and spectroscopic masses are estimated based on the stellar parameters, multi-band magnitudes, distances and extinction values.
Finally, we construct a binary catalog including about 2700 candidates by
analyzing their light curves, fitting the RV data, calculating the binarity parameters from med-resolution spectra, and cross-matching the spatially resolved binary catalog from {\it Gaia} EDR3.
The LAMOST TD survey is expected to get breakthrough in various scientific topics, such as binary system, stellar activity, and stellar pulsation, etc.
}

   \authorrunning{Song Wang et al. }            
   \titlerunning{LAMOST Time-Domain Survey}  
   \maketitle

%
\section{Introduction}           
\label{sect:intro}

Time-domain (hereafter TD) exploration of the sky is at the forefront of modern astronomy.
In recent years, the TD astronomy has rapidly advanced thanks to many wide-field surveys, such as the Palomar Transient Factory \citep[PTF;][]{2009PASP..121.1395L} and Zwicky Transient Facility \citep[ZTF;][]{2019PASP..131a8002B},
the Panoramic Survey Telescope and Rapid Response System  \citep[Pan-STARRS;][]{2004AN....325..636H},
the SkyMapper \citep{2007PASA...24....1K},
the Kepler mission \citep{2010Sci...327..977B},
and the Transiting Exoplanet Survey Satellite \citep[TESS;][]{2015JATIS...1a4003R}.

Most of current TD surveys provide imaging data and focus on the photometrically variable sky, whereas
spectroscopic surveys providing multi-epoch spectra for variable objects are still lacking to date \citep{2018AJ....155....6M}.
The SDSS TD spectroscopic survey, a SDSS-IV eBOSS subproject, is providing repeat observations for about 13000 qusars and 3000 variable stars, including dwarf carbon stars, white dwarf/M dwarf pairs, hypervariable stars, and active ultracool (late-M and early-L) dwarfs \citep{2018AJ....155....6M}.
Recently, LAMOST (Large Sky Area Multi-Object fiber Spectroscopic Telescope; also known as GuoShouJing telescope) started the second 5-year survey program, LAMOST II, containing both non-TD and TD surveys.
In the 5-year survey plan, about 50\% nights (dark/gray nights) are assigned to the low-resolution spectroscopic (LRS; R $\sim$ 1800) survey, and the other 50\% nights (bright/gray nights) to the medium-resolution spectroscopic (MRS; R $\sim$ 7500) survey \citep[see][for more details]{2020arXiv200507210L, 2020ApJS..251...15Z}.

The LAMOST TD survey will monitor about 200 thousand stars with averagely 60 MRS exposures in five years \citep{2020arXiv200507210L}, which provides a great opportunity to get some breakthrough in diverse scientific topics, including binarity, stellar pulsation, star formation, stellar activity, etc.
For example, many attractive binaries are expected to be discovered during their last evolutionary stages, such as wide dwarf-main sequence binaries, symbiotic stars, cataclysmic variables, and even binaries including one neutron star or black hole.
An initial estimation of the precision of the radial velocity (RV) is close to 1 km/s for the MRS data \citep{2020arXiv200507210L}, which is about 3--5 times higher than those obtained from the LRS data \citep{2015RAA....15.1095L}. That means more accurate orbital parameters can be determined for the binaries.
We can also study the variable chromospheric activity of single stars (rotational modulation) or binaries (orbital modulation), by tracing the behavior of the CaII H\&K and H$\alpha$ lines.

In the past few years, Kepler \citep{2010Sci...327..977B} and K2 missions have provided precise TD photometric data for hundreds of thousands of stars, which is a valuable resource for various studies on many topics from exoplanet to asteroseismology.
From 2012 to 2019, LAMOST carried out a LAMOST-Kepler project, using 14 LAMOST plates to almost fully cover the Kepler field of view ($\sim$ 105 square degrees) \citep{2020RAA....20..167F}.
From 2018, the Phase II of the LAMOST-Kepler/K2 survey started aiming at collecting MRS data for more than 50,000 stars located in the Kepler field and six K2 plates \citep{2020ApJS..251...15Z}.
From 2019 to 2020, LAMOST performed a TD survey of four new K2 plates with both LRS and MRS observations.
In Section \ref{ldata.sec}, we describe this project in detail, including data reduction and statistics of the observations and spectra.
We describe the stellar parameter determination and comparison with other databases in Section \ref{par.sec}.
The mass estimation of the sample stars is given in Section \ref{mass.sec}.
In Section \ref{binary.sec}, we present one binary catalog by using different methods.
Finally, we summarize our results and some prospective scientific goals of this project in Section \ref{summary.sec}.

\section{LAMOST OBSERVATION AND DATA REDUCTION}
\label{ldata.sec}

\begin{figure}[!htbp]
   \center
   \includegraphics[width=0.49\textwidth]{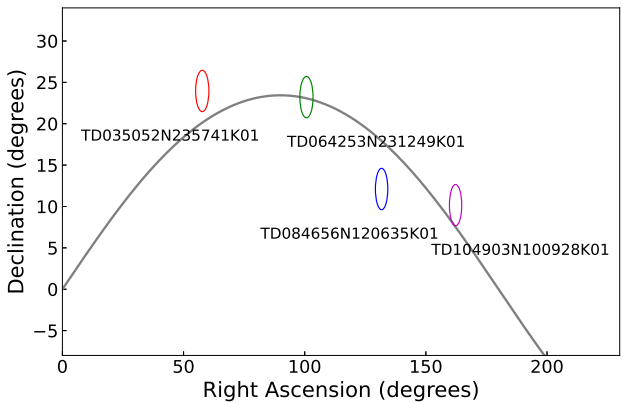}
   \caption{Sky coverage of the four $K$2 plates. The solid line represents the ecliptic plane.}
   \label{sky.fig}
\end{figure}

\begin{figure*}[!htbp]
   \center
   \includegraphics[width=0.49\textwidth]{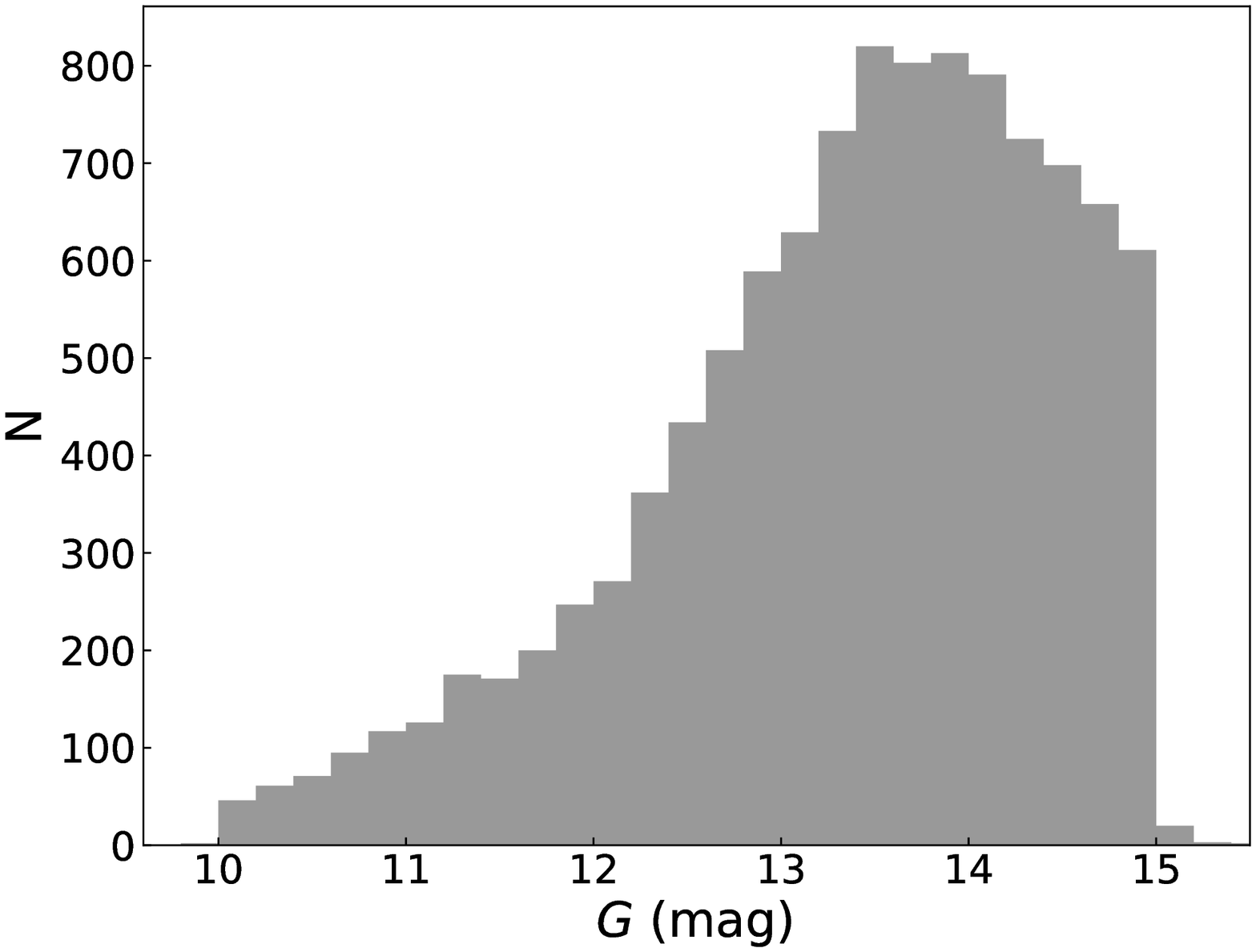}
   \includegraphics[width=0.49\textwidth]{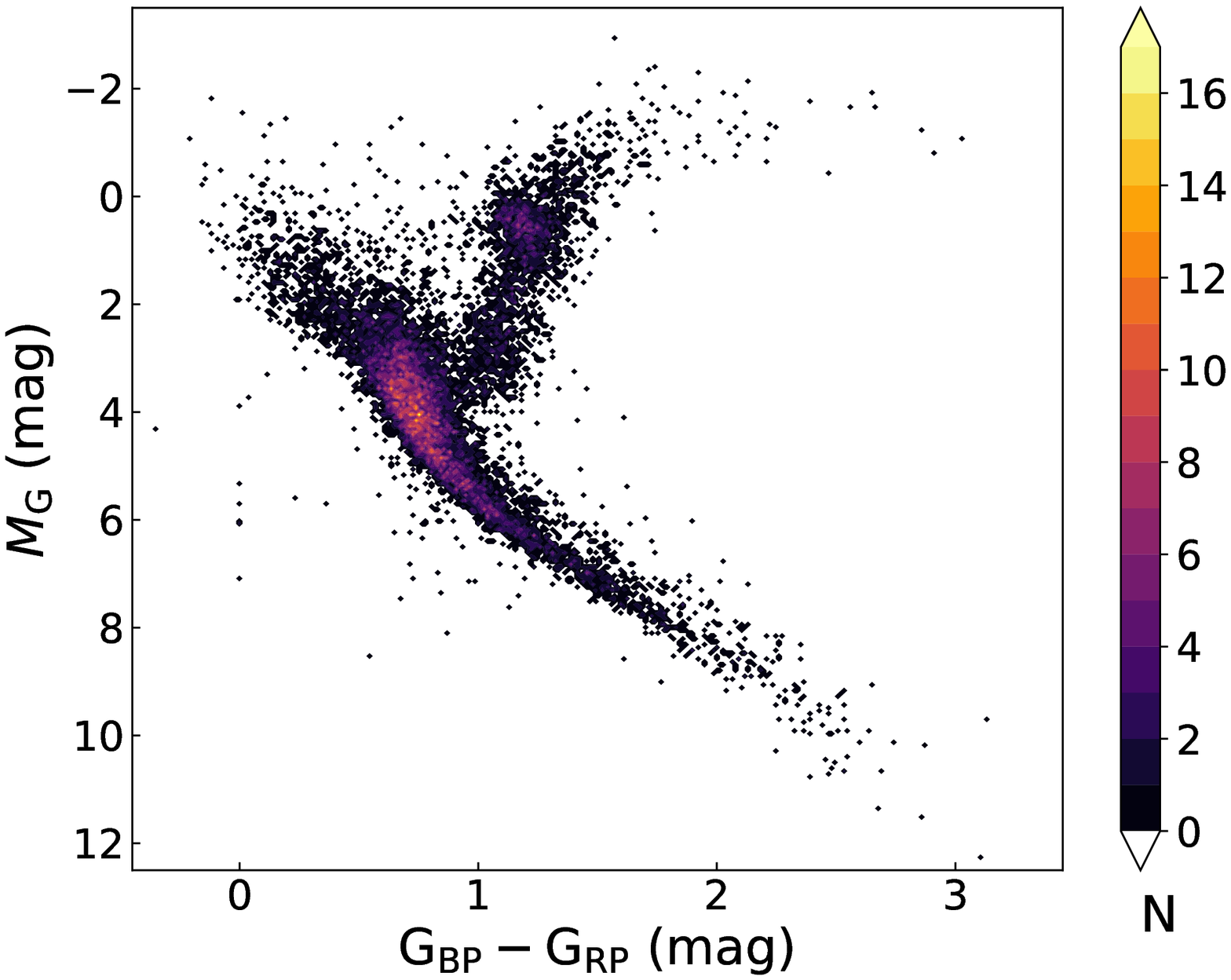}
   \caption{Left Panel: Histogram of the magnitude distribution of our sample stars. The truncation around $G\ =$ 15 mag is due to our selection criteria. Right Panel: Color-magnitude diagram of our sources. The color scale represents the density of stars.}
   \label{hr.fig}
\end{figure*}

This survey includes four footprints in the K2 campaigns (Figure \ref{sky.fig}).
We used the {\it Gaia} DR2 catalog \citep{2018AA...616A...1G} for source selection.
Variable sources recognized by photometric surveys (e.g., ASAS-SN, $K$2) were preferred.
There are totally about 10700 stars in our sample, with magnitudes ranging from $\approx$10 mag to $\approx$15 mag.
Most stars are G- and K-type stars (Figure \ref{hr.fig}).

We performed this survey with both the LRS and MRS observations.
For LRS observation, the wavelength coverage is 3650--9000 \AA\ \citep{2015RAA....15.1095L}.
For MRS observation, the blue and red arms cover wavelength ranges from 4950 \AA\ to 5350 \AA\ and from 6300 \AA\ to 6800 \AA, respectively \citep{2020arXiv200507210L}.
The LRS survey of each plate was observed with 3--10 single 600 s exposures in one observation night; the MRS survey of each plate was observed with 3--8 single 1200 s exposures.
Both the exposure numbers and exposure times may beyond these ranges depending on the observation condition (e.g., seeing).
The fiber assignment contains target stars, flux standard stars, and sky background (Table \ref{four.tab}).

\begin{figure*}[!htbp]
   \center
   \includegraphics[width=0.98\textwidth]{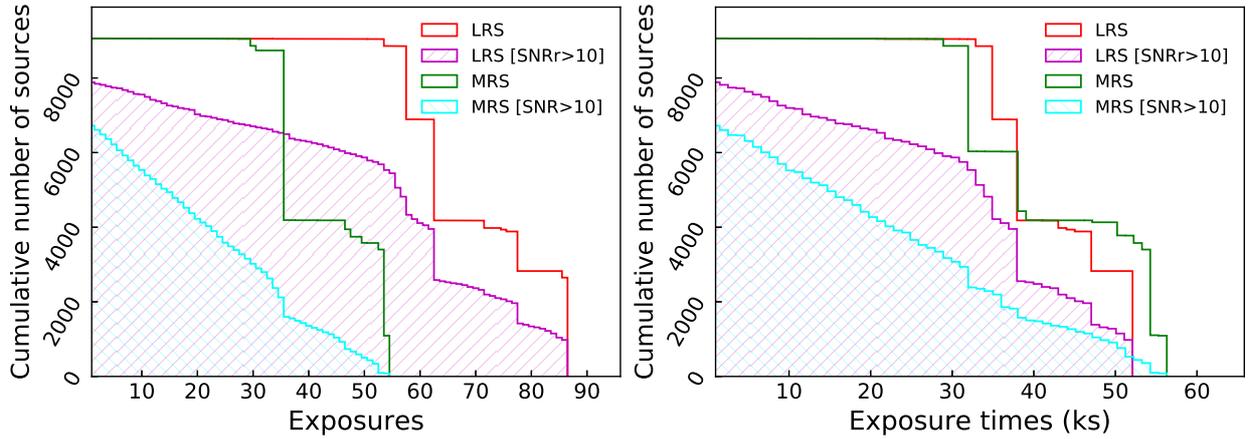}
   \caption{Left Panel: Cumulative histograms of exposure numbers for the LRS and MRS surveys.
   Right Panel: Cumulative histograms of exposure times for the LRS and MRS surveys.}
   \label{exp.fig}
\end{figure*}

From Oct. 2019 to Arp. 2020, the LRS survey was totally performed on 26 dark/gray nights, and the MRS survey was taken on 27 bright/gray nights.
For LRS part, we derived 767,158 and 767,150 spectra in the blue and red arms, respectively, corresponding to a total exposure time of $\approx$46.6 hours.
More than 9000/6800/4100 targets have more than 50/60/70 exposures, and more than 9000/4100/2800 targets were observed more than 30/40/50 ks.
For MRS part, we gained 478,694 spectra for both the blue and red arms, corresponding to an exposure time of $\approx$49.1 hours.
There are more than 8800/4100/3500 targets with more than 30/40/50 exposures, and more than 8800/4100/3700 targets observed more than 30/40/50 ks.
The exposure numbers and exposure times per source are shown in Figure \ref{exp.fig}.

The raw CCD data from the LRS and MRS survey were reduced by the LAMOST 2D pipeline, including bias and dark subtraction, flat filed correction, spectrum extraction, sky background subtraction, wavelength calibration, etc \citep[see][for details]{2015RAA....15.1095L}.
The wavelength calibration of the LRS data was based on the Sr and Th-Ar lamps and night sky lines \citep{2010ApJ...718.1378M}, whereas the wavelength calibration of the MRS data only used the lamps.
A vacuum wavelength scale was applied to the spectra and corrected to the heliocentric frame at last.

In order to show the spectral quality, we calculated the signal-to-noise ratio (SNR) of the $g$-band spectrum for the LRS data and the SNR of the whole spectrum for the MRS data.
We derived 538,760 high-quality spectra (SNR $>$ 10) in the LRS survey, including 479,996, 276,292, and 103,076 spectra with SNR above 20, 50, and 100. They corresponded to a fraction of $\sim$ 89.1\%, 51.3\%, and 19.1\% of the high-quality spectra.
For the MRS survey, we derived 257,558 spectra with SNR above 10, including 176,603, 62,121, and 16,712 ones with SNR higher than 20, 50, and 100, corresponding to a fraction of 68.6\%, 24.1\%, and 6.5\% of the high-quality spectra, respectively.

\begin{figure*}[!htbp]
   \center
   \includegraphics[width=0.98\textwidth]{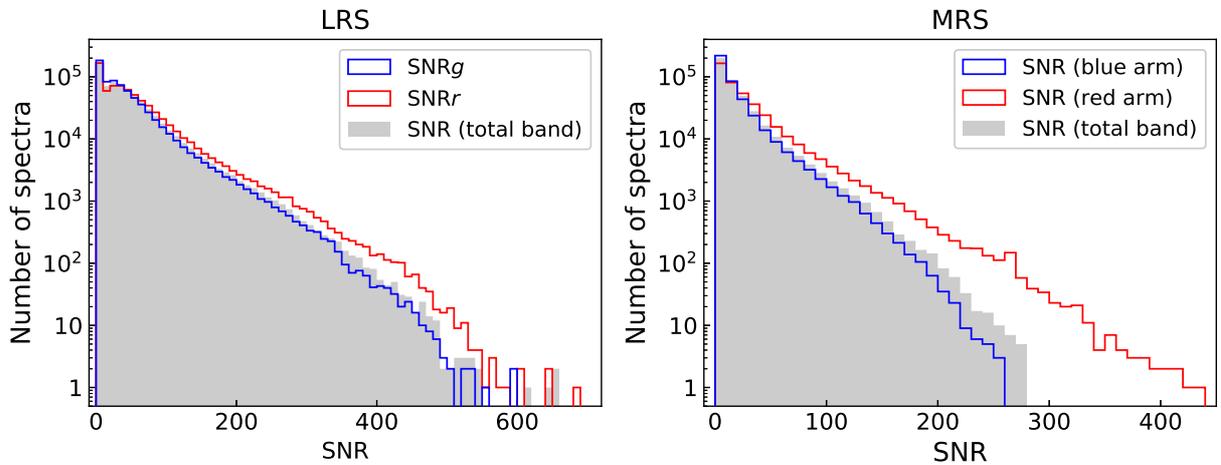}
   \caption{Left Panel: Distribution of SNR for the LRS data. Right Panel: Distribution of SNR for the MRS data.}
   \label{snr.fig}
\end{figure*}

\begin{table*}
\caption[]{Overview of observations of the four $K$2 plates. \label{four.tab}}
\setlength{\tabcolsep}{4.5pt}
 \begin{tabular}{cccccccccc}
  \hline\noalign{\smallskip}
   \multicolumn{6}{c}{} &   \multicolumn{2}{c}{LRS} & \multicolumn{2}{c}{MRS} \\
\cmidrule(lr){7-8} \cmidrule(lr){9-10}
PlanID& R.A.& Dec.&  N$_{\rm star}$& N$_{\rm FS}$ & N$_{\rm sky}$ & N$_{\rm exp,L}$ & N$_{\rm nights,L}$ & N$_{\rm exp,M}$ &  N$_{\rm nights,M}$ \\
  \hline\noalign{\smallskip}
TD035052N235741K01 & 03:50:52.4 & 23:57:41 & 2820 & 78 & 354 &   62 & 8 & 35 &  7\\
TD064253N231249K01  & 06:42:53.9 & 23:12:49 & 2987 & 79 & 316 & 77 & 11 & 54 & 9 \\
TD084656N120635K01 & 08:46:56.0 & 12:06:35 & 2756 & 80 &  503 & 86 & 10 &  53& 11 \\
TD104903N100928K01 & 10:49:03.2 & 10:09:28 & 1885 & 77 & 1253 & 57 & 9 & 35 &  8\\
  \noalign{\smallskip}\hline
\end{tabular}
\smallskip
\tablecomments{\textwidth}{The columns are:
(1) PlanID: the plan name of target field marked with a string of 18 characters;
(2) R.A.: the right ascension of the central star at epoch J2000;
(3) Dec.: the declination of the central star at epoch J2000;
(4) N$_{\rm star}$: the number of input target stars;
(5) N$_{\rm FS}$: the number of flux standard stars;
(6) N$_{\rm sky}$: the number of fibers for sky background measurements;
(7) N$_{\rm exp,L}$: exposure numbers of the LRS survey;
(8) N$_{\rm nights,L}$: observed nights of the LRS survey;
(9) N$_{\rm exp,M}$: exposure numbers of the MRS survey;
(10) N$_{\rm nights,M}$: observed nights of the MRS survey.}
\end{table*}

\section{STELLAR PARAMETER DETERMINATION}
\label{par.sec}

For the spectra obtained in this project, three groups have been using independent approaches to characterize the observed stars and derive stellar parameters.

\subsection{LASP}
\label{lasp.sec}

For both the LRS and MRS data, the LAMOST Stellar Parameter Pipeline \citep[LASP;][]{2015RAA....15.1095L} was used to obtain the atmospheric parameters ($T_\mathrm{eff}$, $\log g$, and [Fe/H]) and RV. It consists of two steps: Correlation Function Initial (CFI) and Ulyss \citep{2011A&A...525A..71W}. The former method provides initial parameter values for Ulyss to determine accurate measurements. The basic idea of the CFI algorithm is based on the template matching method. The synthetic library (from KURUCZ) adopted by the CFI contains 8903 spectra. In general, five best-matching templates are found with nonlinear least-squares minimization method for an observed spectrum. We adopted the linear combination of the stellar parameters of the five templates as initial guesses \citep{2017ApJ...836...77Y} for Ulyss. This method derived all free parameters ($T_\mathrm{eff}$, $\log g$, [Fe/H], and RV) simultaneously via minimizing the squared difference between the observed and the template spectra.

The uncertainties of the parameters can be summarized as 34\,K in $T_{\rm eff}$ , 0.06 \,dex in $\log g$, 0.03\,dex in [Fe/H], and 5.7 km/s in RV for the LRS spectra with SNR $\geq$ 50, and 61\,K in $T_{\rm eff}$ , 0.06 \,dex in $\log g$, 0.04\,dex in [Fe/H], and 1.3 km/s in RV for the MRS spectra with SNR $\geq$ 50. For single epoch spectrum, the errors of the  atmospheric parameters and RV were determined by two factors including the SNR and the best-matched $\chi^2$.
Here we presented a brief description of the estimation of errors, and a more detailed  description is referred to Du B, et al. (2021, in prep).
Basing on a sample of targets having multiple observations, we obtained the precision of the parameters using the following estimator:
\begin{equation} \label{estimator}
 \Delta P_i = \sqrt{N / (N-1)} (P_i - \overline{P})
\end{equation}
where $i$ ($=$ 1, 2, ..., $N$) is one of the individual measurements and $N$ is the total number of measurements for parameter $P$.
Then, we fit both the precision of the parameter and the best-matched $\chi^2$ as functions of the SNR. Through these two functions, the error of the parameter for single epoch spectrum can be calculated according to its SNR and the best-matched $\chi^2$.

Besides the RV determined by LASP, we provided four more RV measurements. They are marked as rv\_ku0, rv\_71el0, rv\_ku1, and rv\_71el1, respectively. The first two RV values were both determined with the cross-correlation method with a set of synthetic spectra as templates. The only difference is that 483 KURUCZ model spectra were selected for rv\_ku0 and 71 spectra from ELODIE library for rv\_71el0. The latter two values were further calibrated with RV zero point (RVZP) derived by Th-Ar and Sc arc lamps. A brief description of the cross-correlation method was presented as follows. First, a rough RV value was derived by matching an observed spectrum with templates shifted from $-$600 km/s to 600 km/s in step of 40 km/s. Second, matches were carried out between the observed spectrum and templates shifted from $-$60 km/s to 60 km/s in step of 1 km/s. Finally, the RV was determined from the highest peak of a group of correlation functions. More details are referred to \citet{2019ApJS..244...27W}.
These RV values are not used in following analysis.

\subsection{The DD-Payne}
\label{dd.sec}


For the LRS data, we have also determined the stellar parameters with the DD-Payne \citep{Ting2019,Xiang2019}. The DD-Payne derives the stellar parameters with a hybrid method that combines the data-driven approach with priors of astrophysical modeling \citep{Ting2017, Xiang2019}, utilizing the neural-network spectral interpolating and fitting algorithm of the Payne \citep{Ting2019}. We inherited the LAMOST DD-Payne model of \citet{Xiang2019}, which constructs a neural-net spectral model using the LAMOST spectra that have accurate stellar parameters from high-resolution spectra from GALAH DR2 \citep{Buder2018} and the value-added stellar parameter catalog of the APOGEE DR14 derived with the Payne \citep{Ting2019}. The DD-Payne delivers $T_{\rm eff}$, $\log g$, and elemental abundances for 16 elements, C, N, O, Na, Mg, Al, Si, Ca, Ti, Cr, Mn, Fe, Co, Ni, Cu, Ba, as well as their error estimates from single-epoch spectra. The error estimates are obtained by propagating the spectral flux uncertainties in the fitting. To yield statistically realistic error estimates, \citet{Xiang2019} further scaled the fitting errors to the dispersion of repeat observations. For a spectrum with SNR above 50, typical aleatoric uncertainty of the parameter estimates is 30\,K in $T_{\rm eff}$, 0.07\,dex in $\log g$, 0.03--0.1\,dex in the elemental abundance [X/H], except for that [Cu/H] and [Ba/H] exhibit larger uncertainties (0.2--0.3\,dex).

\begin{figure}[!htbp]
   \center
   \includegraphics[width=0.49\textwidth]{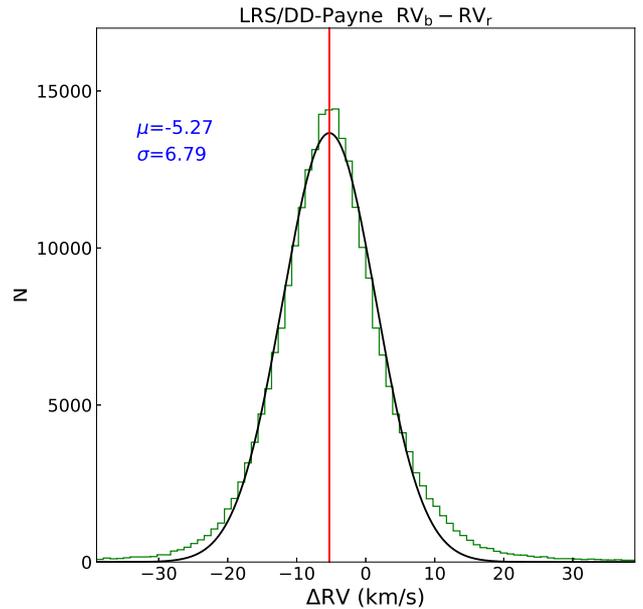}
   \caption{Comparison of the RV values derived with the DD-Payne method from the blue and red bands of the LRS data.}
   \label{xiangrv.fig}
\end{figure}

The DD-Payne model of \citet{Xiang2019} is built on spectra in rest frame while itself does not deliver the stellar RV values. We determined RV with a cross-correlation algorithm, similar to that of the LSP3 \citep{Xiang2015}. We adopted the PHOENIX synthetic spectra \citep{Husser2013}, after degrading to the LAMOST line spread function, as the templates of RV determination. Besides the RV derived from the full LAMOST spectra (3800--9000 \AA), we also delivered the RV$_{\rm b}$ and RV$_{\rm r}$ from the blue- and the red-arm spectra separately, as it is found that there is considerable systematic offset in the wavelength calibration between the blue- and red arms of the LAMOST spectrographs (Fig.\,\ref{xiangrv.fig}). This systemic offset has been reported by \citet{2019ApJS..240...10D} that the RV value calculated with the $H_{\alpha}$ line in the red arm is higher by $\sim$ 7 km/s than those from the blue arm.
In following analysis we used the RV value from the blue arm.


\subsection{SLAM}
\label{slam.sec}

For the MRS data, we also derived the stellar parameters (e.g., $T_{\rm eff}$, log$g$, and [Fe/H]) with the Stellar LAbel Machine (SLAM) \citep{2020ApJS..246....9Z, 2020RAA....20...51Z}, which is a machine learning method like DD-Payne but based on support vector regression (SVR).
SLAM can generally determine stellar labels over a wide range of spectral types.
It consists of three steps, including data pre-processing (i.e., spectra normalization and training data standardization),  SVR model training for each wavelength pixel, and stellar label predicting for observed spectra.
Previous tests on the LAMOST MRS data showed that for a spectrum with $\mathrm{SNR}\approx50$ the precisions of $T_{\rm eff}$, log$g$, and [Fe/H] are about 65 K, 0.02 dex, and 0.06 dex, respectively \citep{2020RAA....20...51Z}.

RVs of spectra were first estimated with a cross-correlation function maximization method\footnote{\url{https://github.com/hypergravity/laspec}.} \citep{2021arXiv210511624Z} and were used to shift the normalized spectra to the same scale.
Then SLAM was trained on the synthetic spectral grid based on ATLAS9 model \citep{2018A&A...618A..25A} which is degraded to $R\sim7500$, and was used to derive stellar labels including $T_\mathrm{eff}$, $\log{g}$, [Fe/H] and [$\alpha$/Fe].

\begin{figure}[!htbp]
\vspace{0.1in}
   \center
   \includegraphics[width=0.49\textwidth]{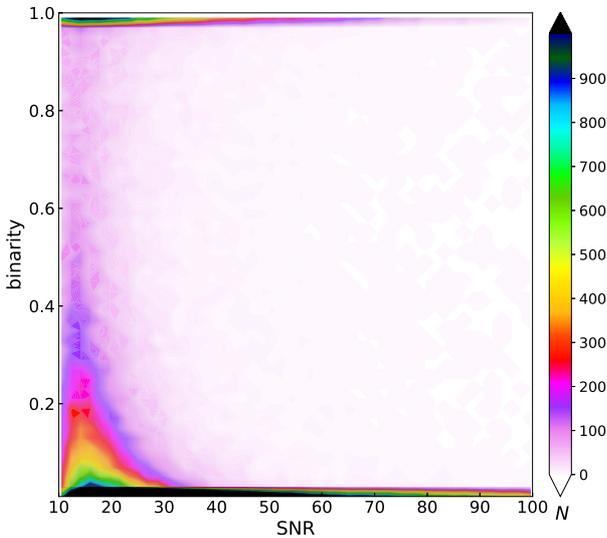}
   \caption{Distribution of the binarity parameter for the MRS spectra with SNR $>$ 10. The color scale represents the density of stars.}
   \label{binarity.fig}
\end{figure}

To efficiently cope with spectroscopic binaries, we estimated a ``binarity" parameter for each spectrum.
We generated 100,000 spectra for single stars and 100,000 for binaries basing on stellar evolutionary model \citep{2016ApJ...823..102C, 2016ApJS..222....8D} and the synthetic spectral grid \citep{2018A&A...618A..25A}, trained a convolutional neural network (CNN) as a classifier, and finally predicted the binarity values of observed spectra.
This method is initially described in Jing et al. (2021, in prep) and applied to the LAMOST LRS spectra.
Figure \ref{binarity.fig} shows the distribution of the binarity parameter for the MRS spectra with SNR $>$ 10. The subpopulation with binarity $>$ 0.9 are mostly double-lined spectroscopic binaries.
Manual inspection of those spectra shows that this classification method is very efficient.
Currently this method is still being improved and tested on more LAMOST MRS spectra (Zhang et al. 2021, in prep).

\subsection{Comparison between different methods}

As shown above, we used three independent methods to determine the stellar parameters.
Since the LASP method was used to derive the parameters for both the LRS and MRS data, we compared their results with those from DD-Payne (for LRS data) and SLAM (for MRS data).
The spectra with SNR $>$ 50 and 4000 K $<$ $T_{\rm eff}$ $<$ 7500 K (LASP results) were used for comparison.

\begin{figure*}[!htbp]
   \center
   \includegraphics[width=0.98\textwidth]{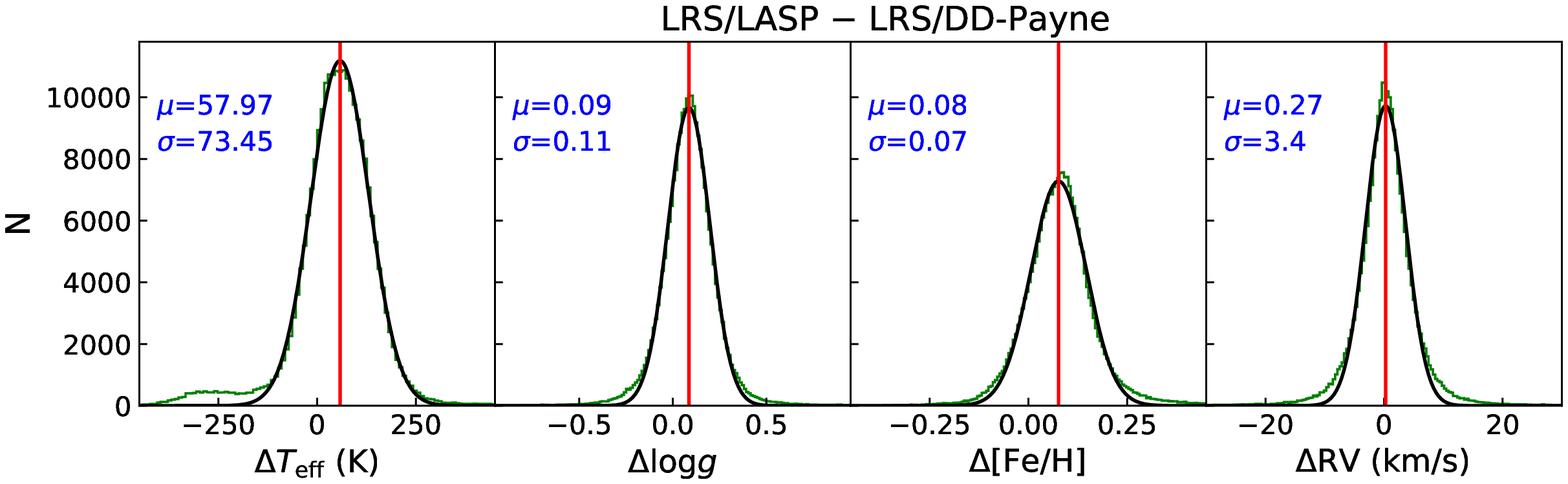}
   \includegraphics[width=0.98\textwidth]{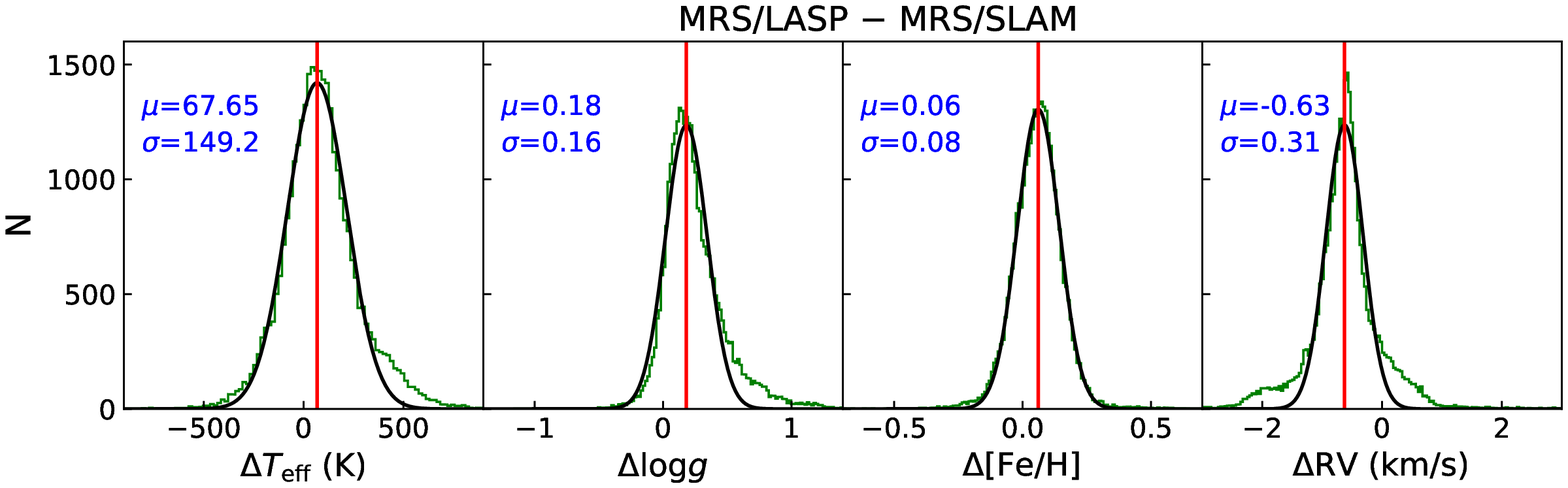}
   \caption{Top panel: Comparison of the $T_\mathrm{eff}$, $\log g$, [Fe/H], and RV values between LASP and DD-Payne using the LRS data . The black lines are the best fitting with a single Gaussian distribution to the histograms (green). Bottom panel: Comparison of the $T_\mathrm{eff}$, $\log g$, [Fe/H], and RV values between LASP and SLAM using the MRS data. }
   \label{hx.fig}
\end{figure*}

\begin{figure}[htp!!!]

   \includegraphics[width=0.49\textwidth]{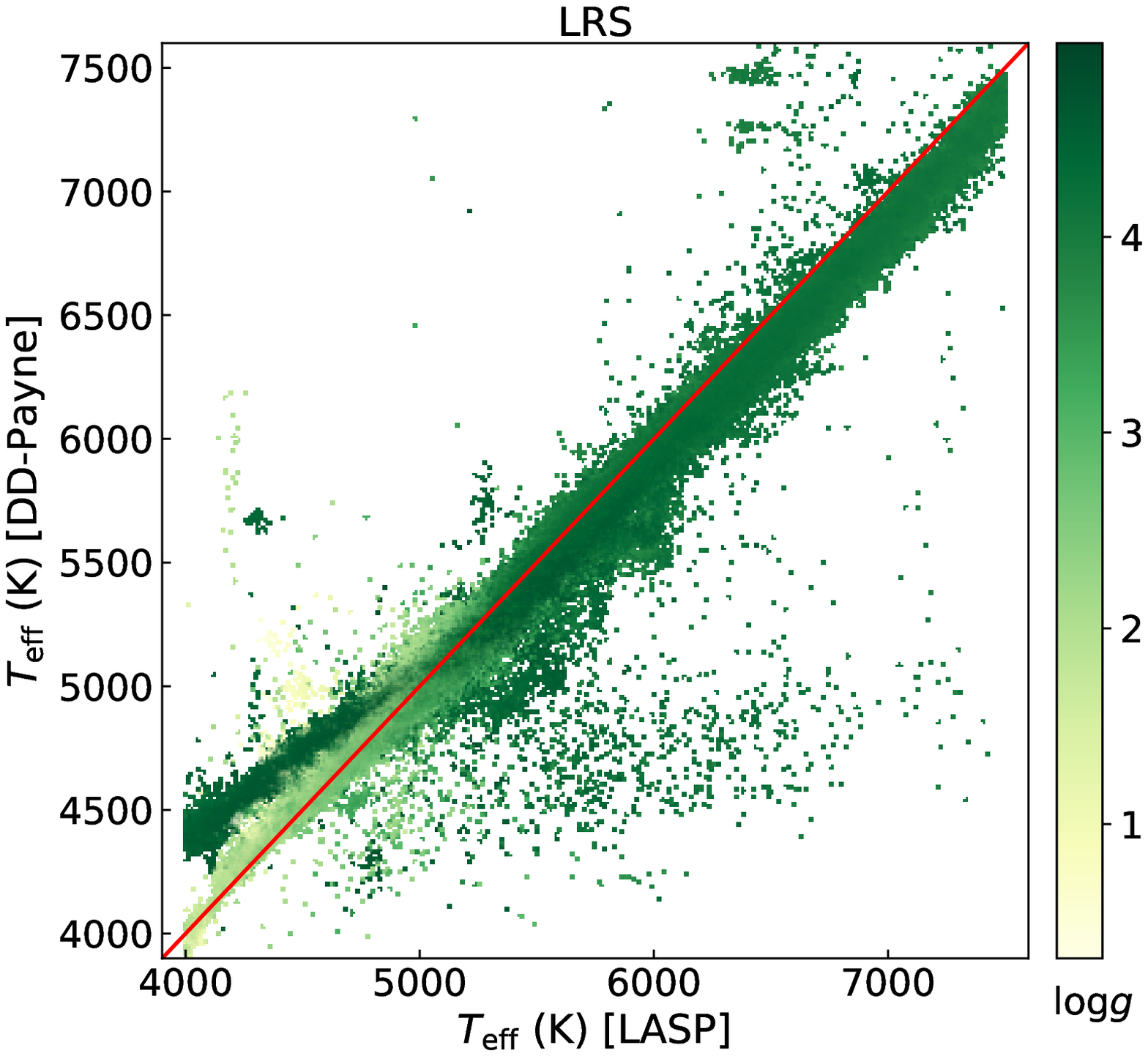}\\
   \includegraphics[width=0.49\textwidth]{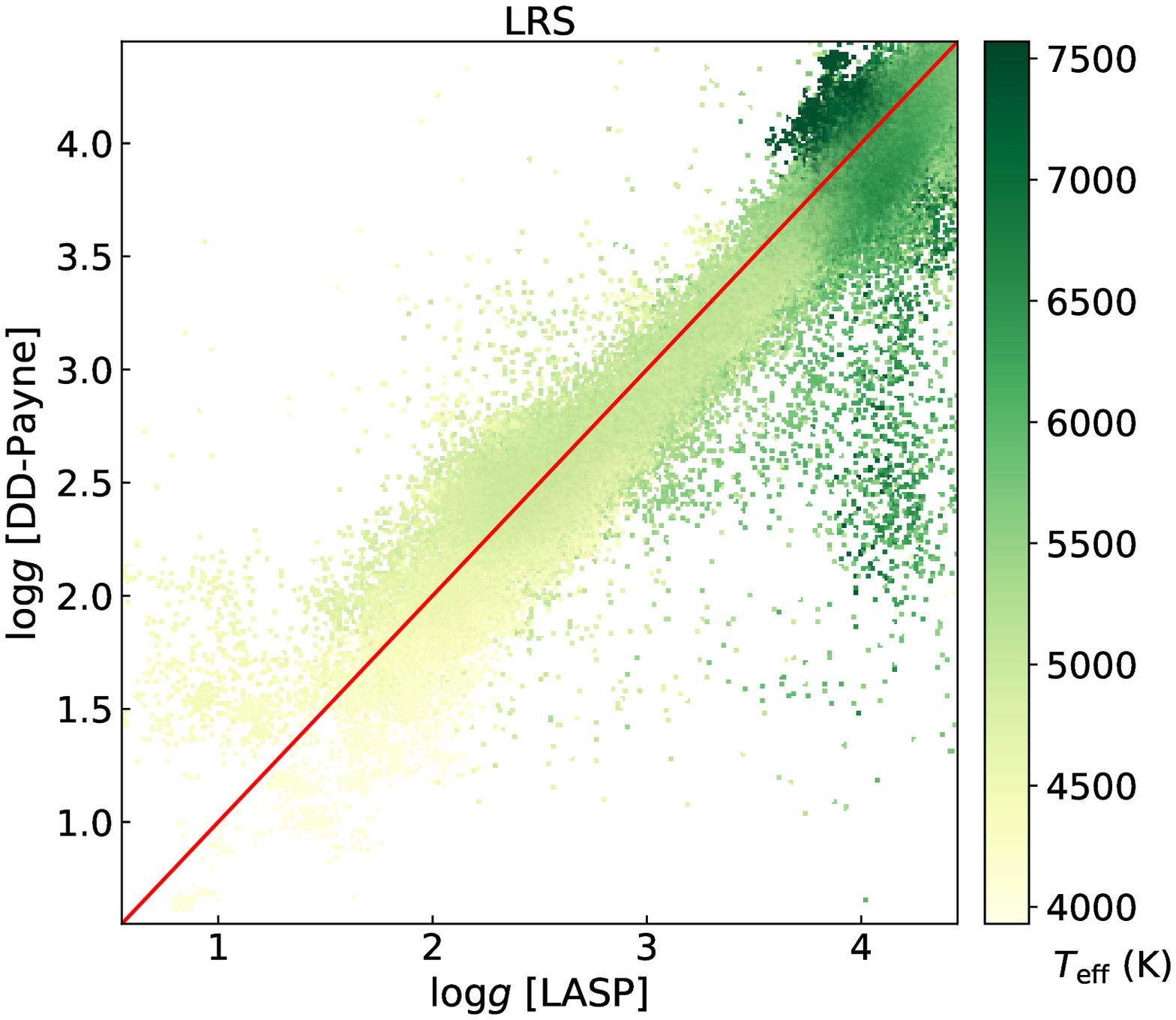}\\
   \includegraphics[width=0.49\textwidth]{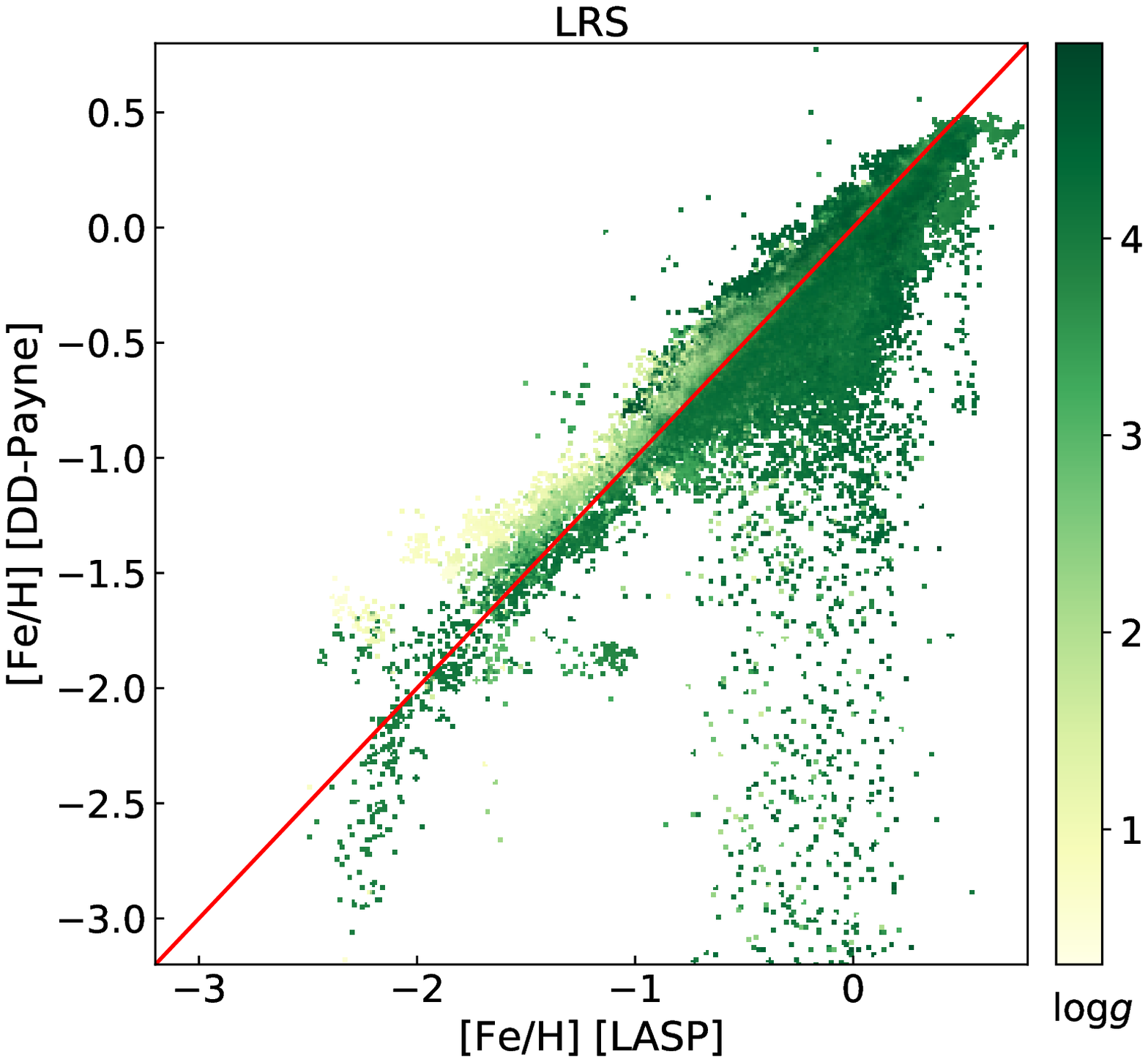}\\
   \caption[]{Top Panel: Comparison of $T_\mathrm{eff}$ between LASP and DD-Payne using the LRS data. The colorbar represents $\log g$. 
   Middle Panel: Comparison of $\log g$ between LASP and DD-Payne. The colorbar represents $T_\mathrm{eff}$.
   Bottom Panel: Comparison of [Fe/H] between LASP and DD-Payne. The colorbar represents $\log g$.}
   \label{tefflog1.fig}
\end{figure}

\begin{figure}
      \includegraphics[width=0.49\textwidth]{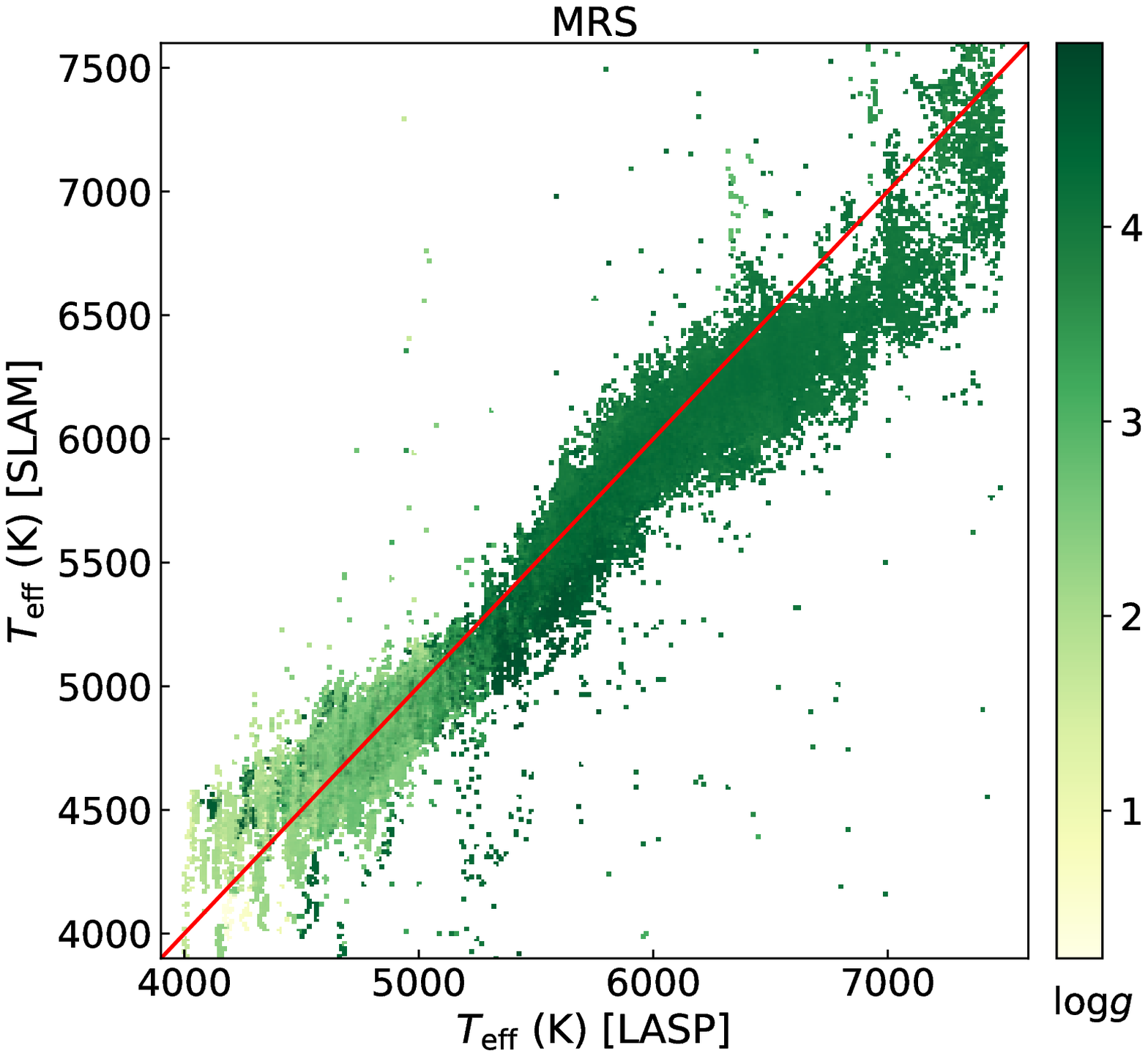}\\
   \includegraphics[width=0.49\textwidth]{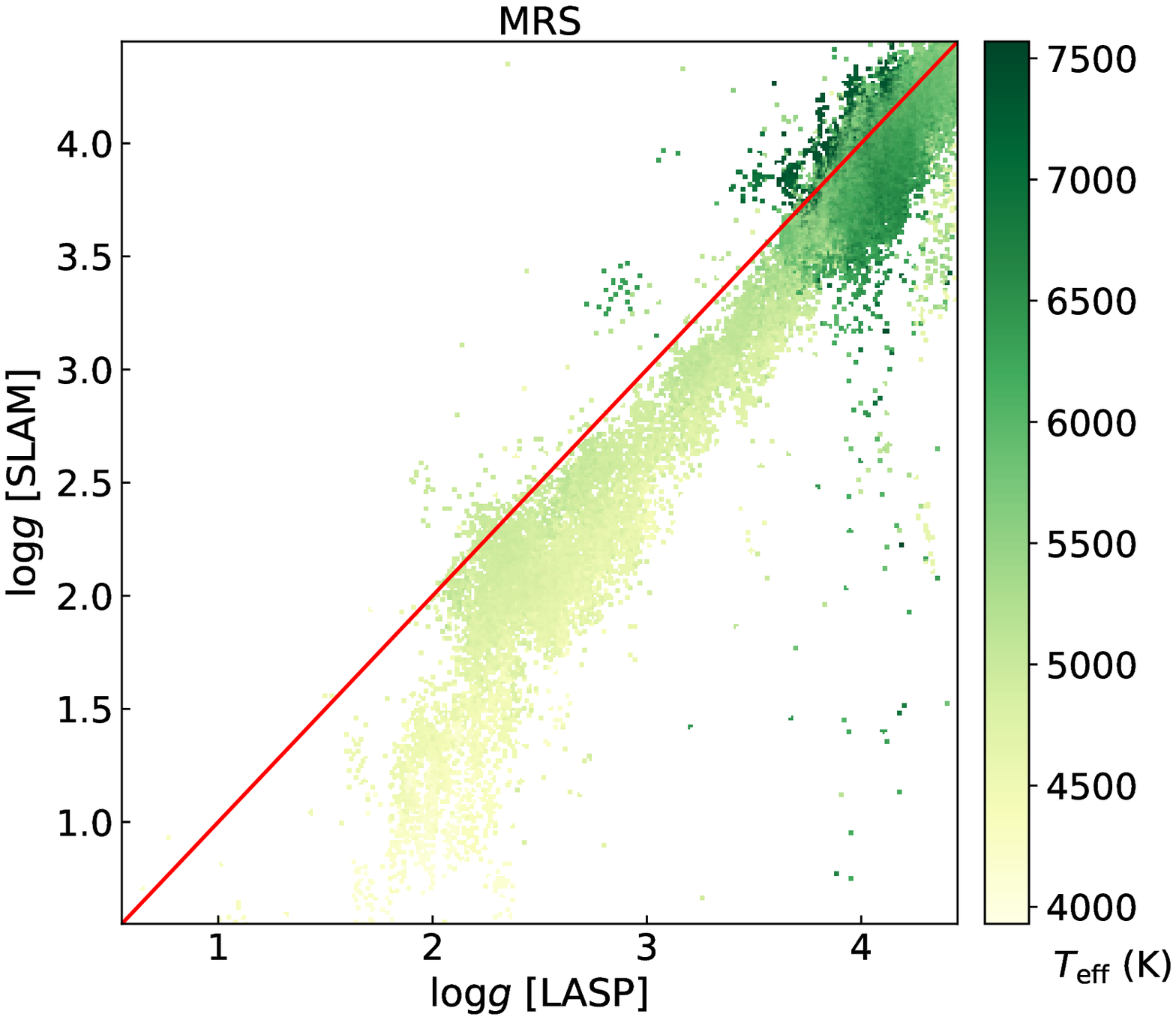}\\
      \includegraphics[width=0.49\textwidth]{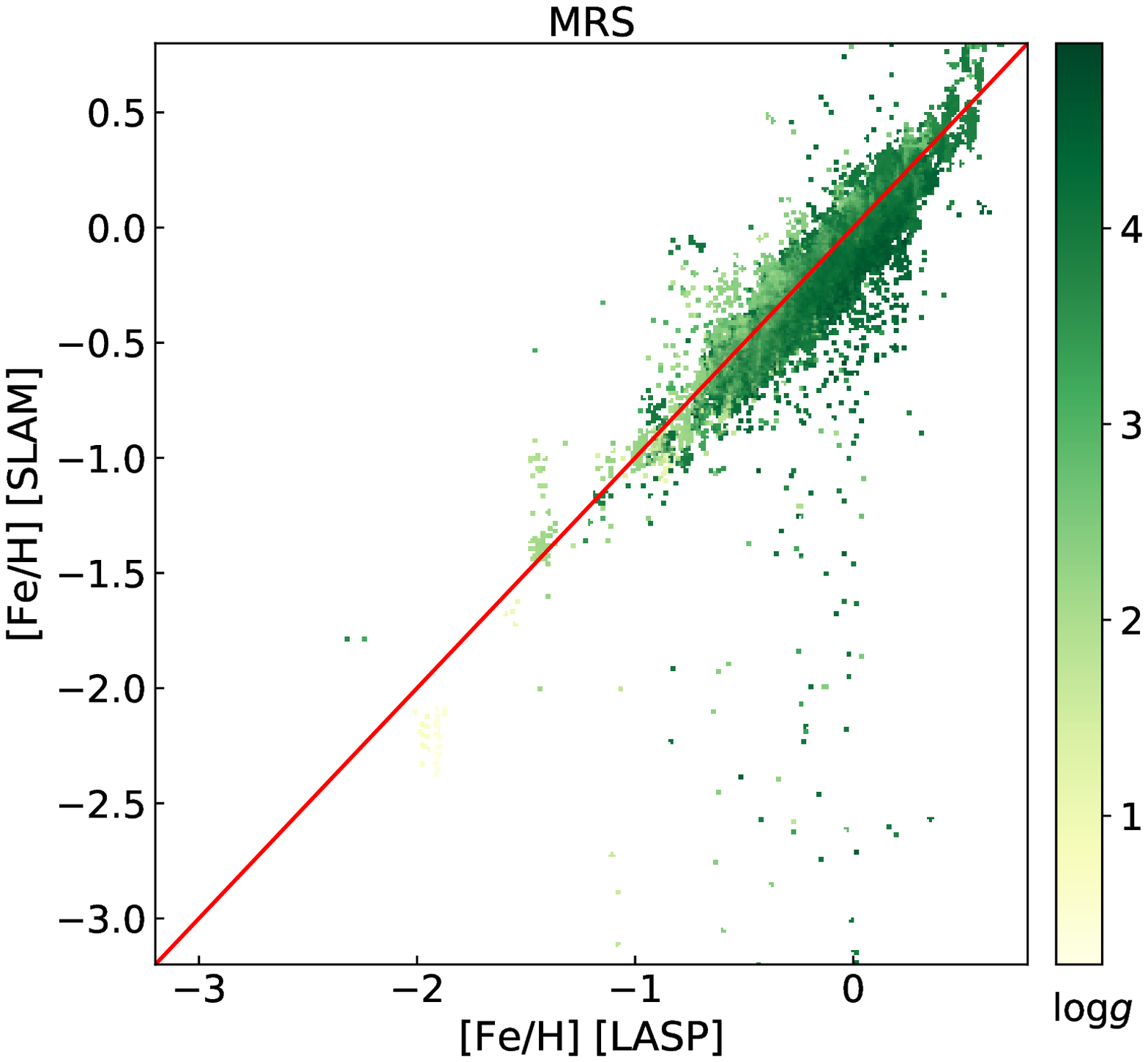}\\
   \caption{Top Panel: Comparison of $T_\mathrm{eff}$ between LASP and SLAM using the MRS data. The colorbar represents $\log g$. 
   Middle Panel: Comparison of $\log g$ between LASP and SLAM. The colorbar represents $T_\mathrm{eff}$.
   Bottom Panel: Comparison of [Fe/H] between LASP and SLAM. The colorbar represents $\log g$.}
   \label{tefflog2.fig}
\end{figure}

In general, most of the parameters obtained from different methods are in good agreement (Figure \ref{hx.fig}).
There are some objects showing lower effective temperatures ($\approx$ 250 K) from LASP results than those from DD-Payne (Figure \ref{tefflog1.fig}).
These objects are mostly cool dwarfs, which have temperature estimations ranging from $\sim$4000 K to $\sim$4700 K in LASP results but ranging from $\sim$4300 K to $\sim$4900 K in DD-Payne results.
A group of objects classified as dwarfs by LASP (log$g$ $\gtrsim$ 3.5) have log$g$ estimations by DD-Payne lower than 3.0.
For the MRS data, some hot dwarfs ($T_{\rm eff} \gtrsim$ 6500 K from the LASP results) show higher temperatures (around 500 K) than those from SLAM (Figure \ref{tefflog2.fig}).
The surface gravity shows deviation from a symmetric gaussian distribution. Most of these objects are cool dwarfs ($T_{\rm eff}$ $\lesssim$ 4500 K).
This systematic offset is mainly caused by the different training set: LASP uses the empirical template library ELODIE, while SLAM uses the synthetic spectral grid from ATLAS9 model.

\subsection{RV correction}
\label{rvcor.sec}

Due to the temporal variation of the zero-points, small systemic offsets exist in RV measurements (\citealp{2019RAA....19...75L,2020ApJS..251...15Z,2021arXiv210511624Z}).
Therefore, the RV value of each spectrum (i.e., each fiber at each exposure) needs a correction with corresponding zero point.
For the MRS data, both the LAMOST pipeline and \citet{2019ApJS..244...27W} determined a universal RVZP for each spectrograph by comparing the measured RVs to those of RV standard stars selected from APOGEE data \citep{2018AJ....156...90H}.
This only corrects the systemic RVZP offsets between different spectrographs.
\citet{2019RAA....19...75L} proposed a method to correct the temporal RVZP variation by using ``RV-constant" stars in each spectrograph.
However, we found that there are only few ``RV-constant" stars for some spectrographs in the observations of one field. If the RVZP varies abruptly in one observation, these ``RV-constant" stars will be excluded, or this observation has to be abandoned.

\begin{figure*}[!htbp]
   \center
   \includegraphics[width=0.98\textwidth]{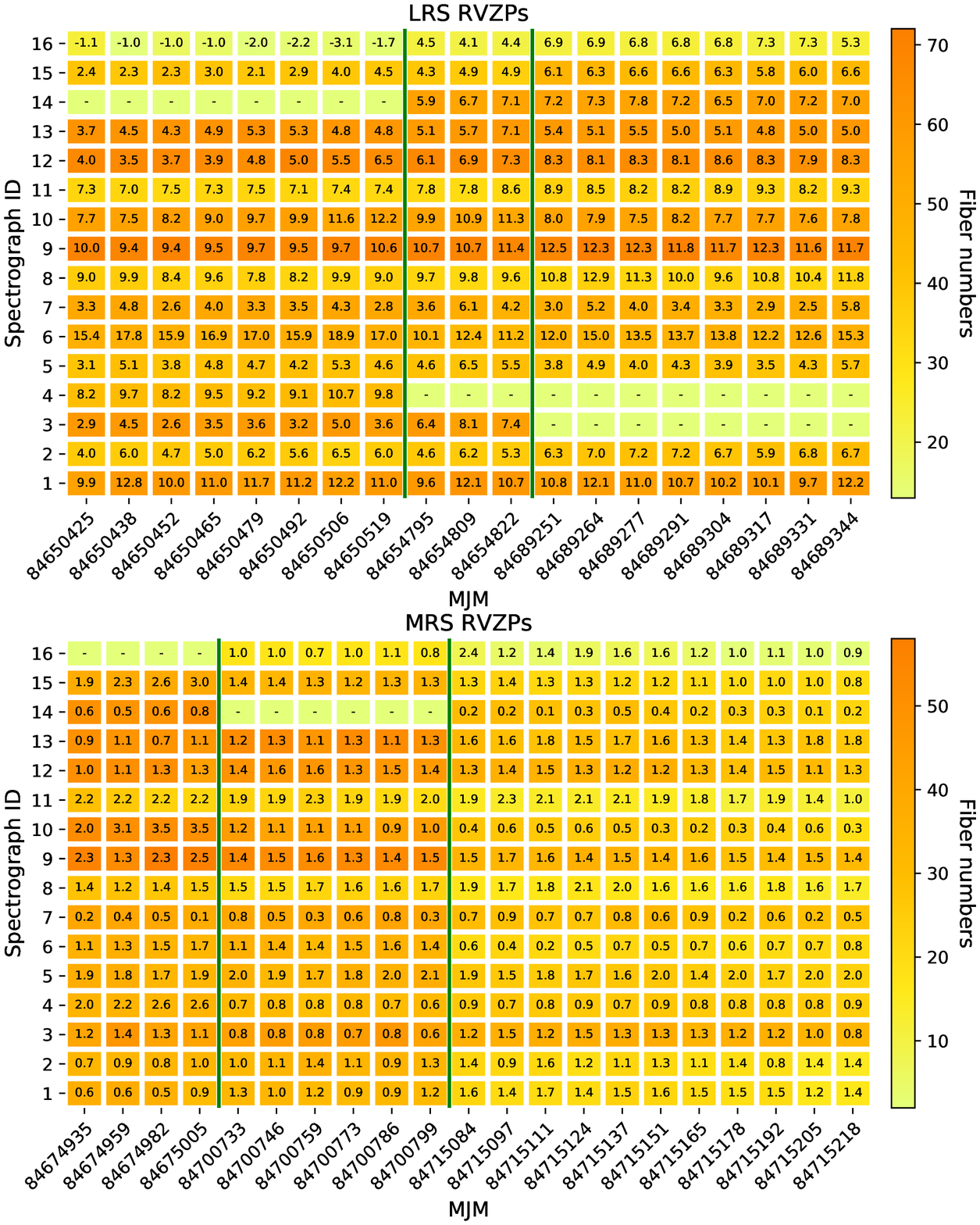}
   \caption{Top panel: Example of distribution of the RVZPs as a 2D-function of the spectrograph ID and observed epochs (local MJM) for the TD035052N235741K01 plate by using the LRS data. The numbers mean the RVZP values in km/s, and the color represents common stars used to calculated the offset values. The vertical green lines split different nights.
   Bottom panel: Example of distribution of the RVZPs as a 2D-function of the spectrograph ID and observed epochs (local MJM) for the TD035052N235741K01 plate by using the MRS data.}
   \label{vcor.fig}
\end{figure*}

Here, we used the Gaia DR2 data to determine the RVZPs for each spectrograph exposure by exposure, and applied them as the common RV shift of the fibers in the same spectrograph.
For each spectrograph, we compared the RVs of the common objects in each exposure and those from {\it Gaia} DR2, and determined a median offset $\Delta RV$ with two or three iterations. One can determine the RVZP-corrected RVs by adding the offset $\Delta RV$, and use them to compare with external RV databases (e.g., APOGEE). As an example, Figure \ref{vcor.fig} shows the calculated RVZPs (i.e., $\Delta RV$) of each spectrograph in some exposures of the TD035052N235741K01 plate.

\subsection{Weighted average values of the stellar parameters}

\begin{figure}[!htbp]
   \center
   \includegraphics[width=0.49\textwidth]{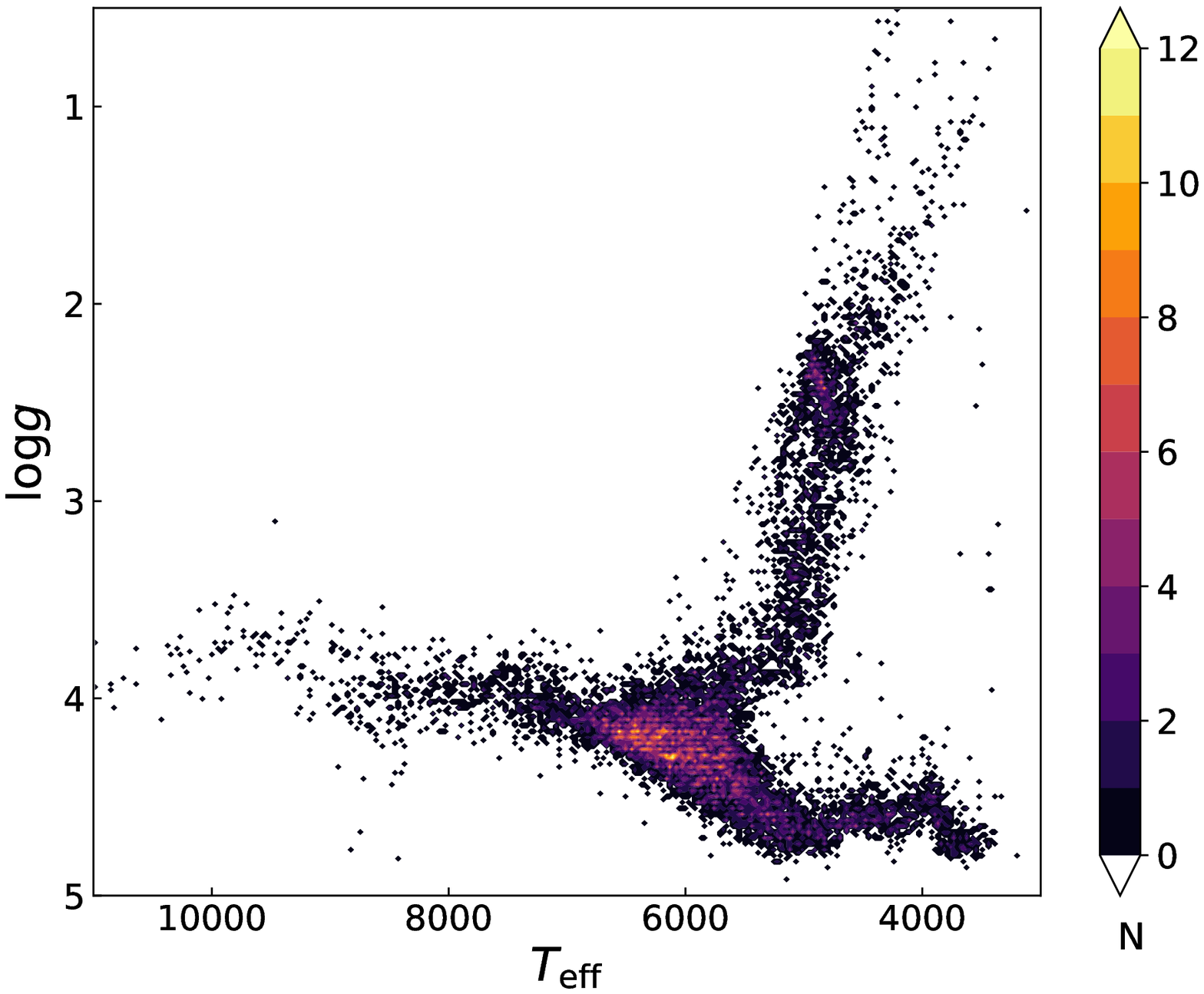}
   \caption{Hertzsprung-Russel diagram of the sample stars. The color scale represents the density.} 
   \label{tefflogg.fig}
\end{figure}

Most of these targets were observed at multiple epochs, which means we can obtain average values of the stellar parameters and RV for each target.
By using the spectra with SNR above 10, We derived SNR-weighted average values and corresponding errors for the stellar parameters of each target with the formulae \citep{2020ApJS..251...15Z}:
\begin{equation} \label{eqweight}
\overline{P} = \frac{\sum_k w_k \cdot P_{k}}{\sum_k w_k}
\end{equation}
and
\begin{equation}
\sigma_w(\overline{P}) = \sqrt{\frac{N}{N-1}\frac{\sum_k w_k \cdot (P_{k} - \overline{P})^2}{\sum_k w_k}}.
\end{equation}
The index $k$ is the epoch of the measurements of parameter $P$ (i.e., $T_{\rm eff}$, log$g$, [Fe/H], and RV) for each star, and the weight $w_k$ is estimated with the square of the SNR of each spectrum. Figure \ref{tefflogg.fig} shows the distribution of our samples in the log$g$--$T_{\rm eff}$ diagram.

\begin{figure*}[!htbp]
   \center
   \includegraphics[width=0.98\textwidth]{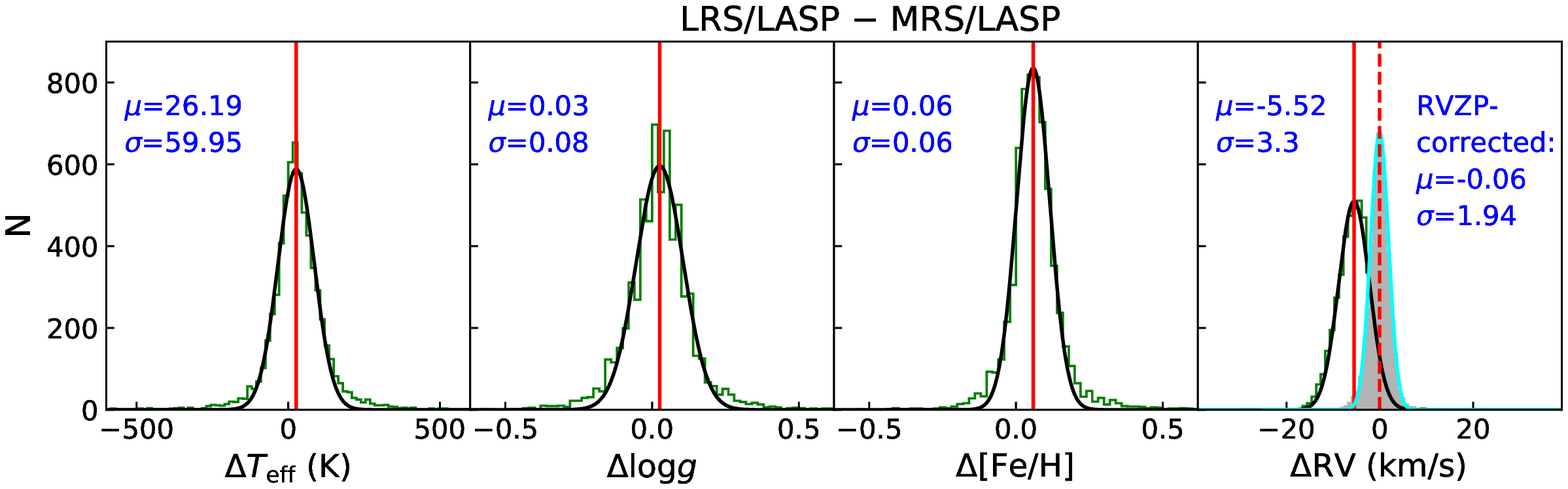}
   \caption{Comparison of the $T_\mathrm{eff}$, $\log g$, [Fe/H], and RV values from LRS and MRS data by using the LASP method. The black lines are the best fitting with a single Gaussian distribution to the histograms (green) The shaded histogram represent the difference of the RVZP-corrected RVs from the LRS and MRS data.}
   \label{comLM.fig}
\end{figure*}

We used the weighted average values (from LASP estimation) to make a comparison of the parameters derived from the LRS and MRS data.
Generally, the values of $T_\mathrm{eff}$, $\log g$, and [Fe/H] from LRS are in good agreement with those from MRS (Figure \ref{comLM.fig}).
There is a systematic offset between the LRS and MRS RV measurements ($-5.52\pm3.30$ km/s).
After correcting the RVZP (Section \ref{rvcor.sec}), the offset reduces to $-$0.06$\pm$1.94 km/s.
The systematic offset nearly disappears, suggesting that our RV correction method is reasonable and valid.

\subsection{Comparison with APOGEE}

\begin{figure*}[!htbp]
   \center
   \includegraphics[width=0.98\textwidth]{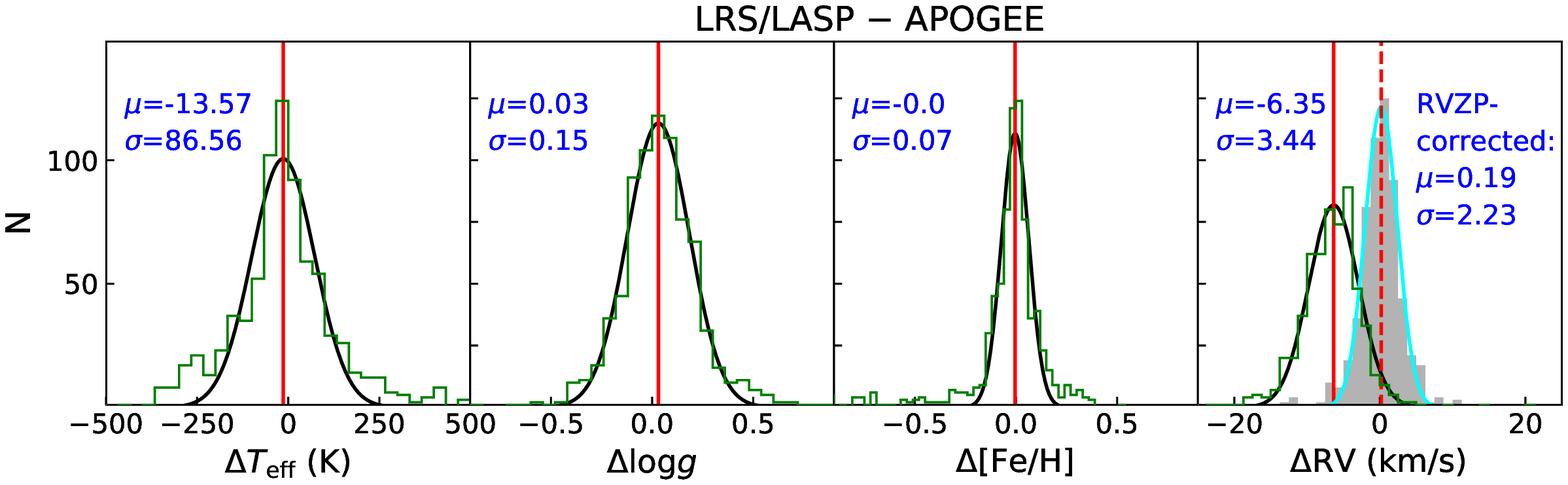}
      \includegraphics[width=0.98\textwidth]{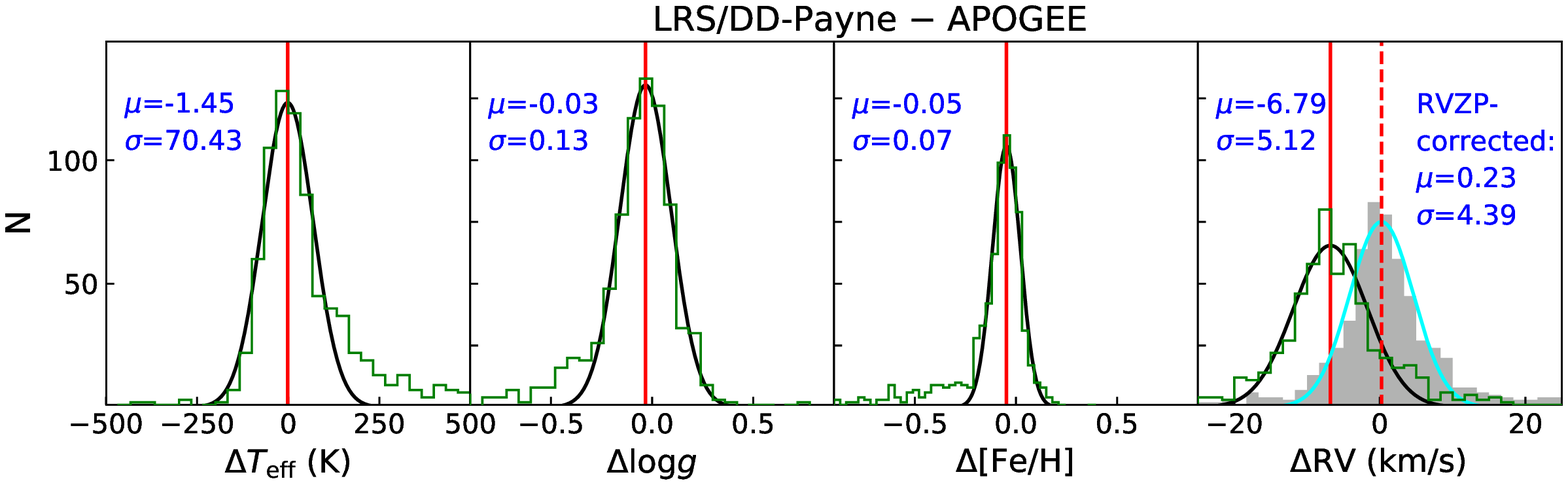}
   \includegraphics[width=0.98\textwidth]{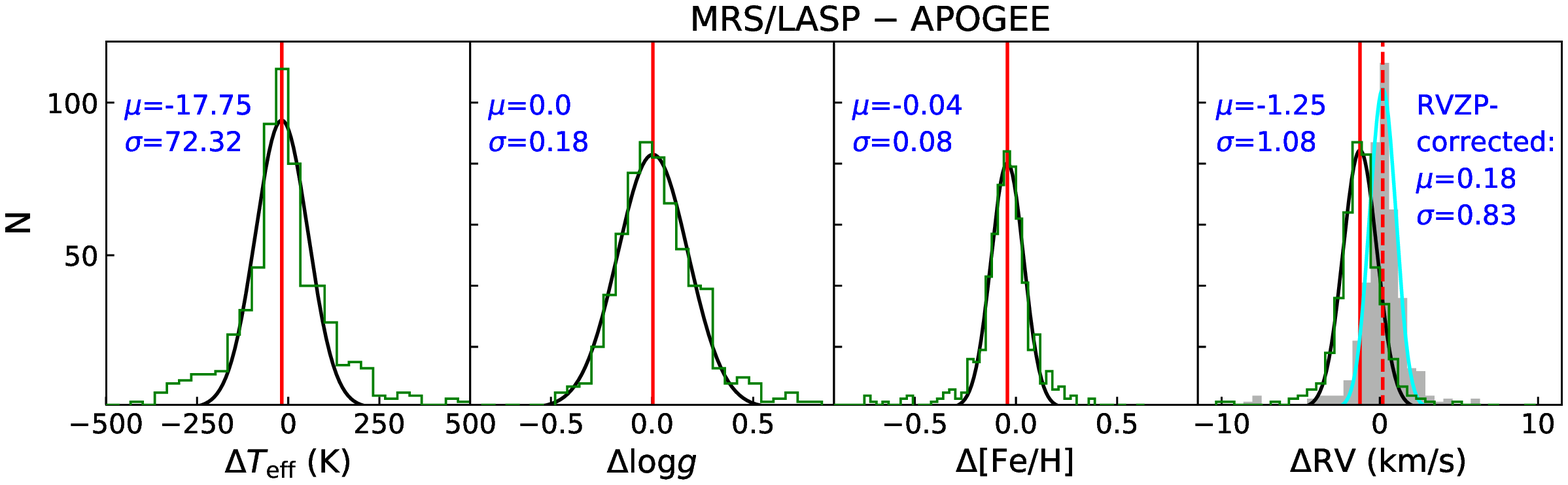}
      \includegraphics[width=0.98\textwidth]{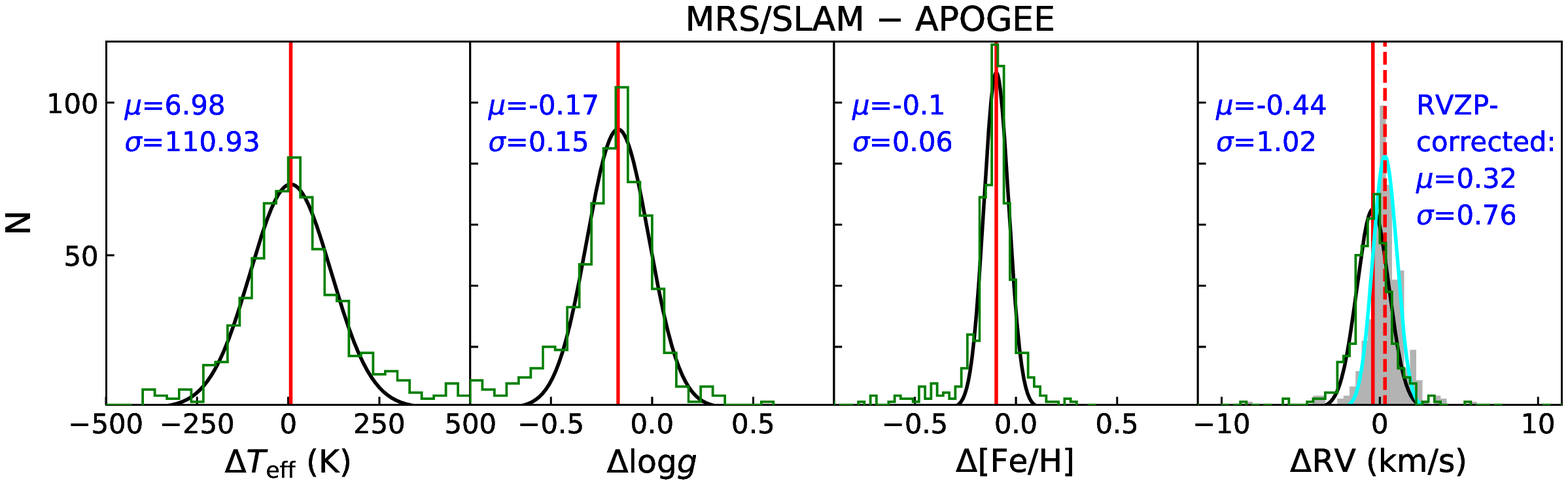}
   \caption{Comparison of the $T_\mathrm{eff}$, $\log g$, [Fe/H], and RV values from this study and APOGEE data. The method and data from top to bottom are: LASP using LRS data, DD-Payne using LRS data (the RV determination from only the blue-arm spectra is adopted), LASP using MRS data, and SLAM using MRS data. The shaded histogram represent the difference of the RVZP-corrected RVs and APOGEE.}
   \label{comLA.fig}
\end{figure*}

We cross-matched our sample with the APOGEE DR16 catalog, and there are 1,001 common stars.
In general, the values of $T_\mathrm{eff}$, $\log g$ and [Fe/H] from both the LRS and MRS surveys and those from APOGEE are consistent (Figure \ref{comLA.fig}).
It can be seen that the RVZP-corrected RVs show good agreement with those of APOGEE.
As noted in Section \ref{dd.sec} that the LRS RVs from the blue and red arms show a systemic offset of $\approx$5--7 km/s, we found the LRS RVs from red arm agree well with those of APOGEE, with very small offset ($\mu$ = $-$0.91 km/s; $\sigma$ = 3.48 km/s), although the RVZP-corrected values show little improvement ($\mu$ = $-$0.76 km/s; $\sigma$ = 3.06 km/s).

\begin{figure*}[!htbp]
   \center
   \includegraphics[width=0.98\textwidth]{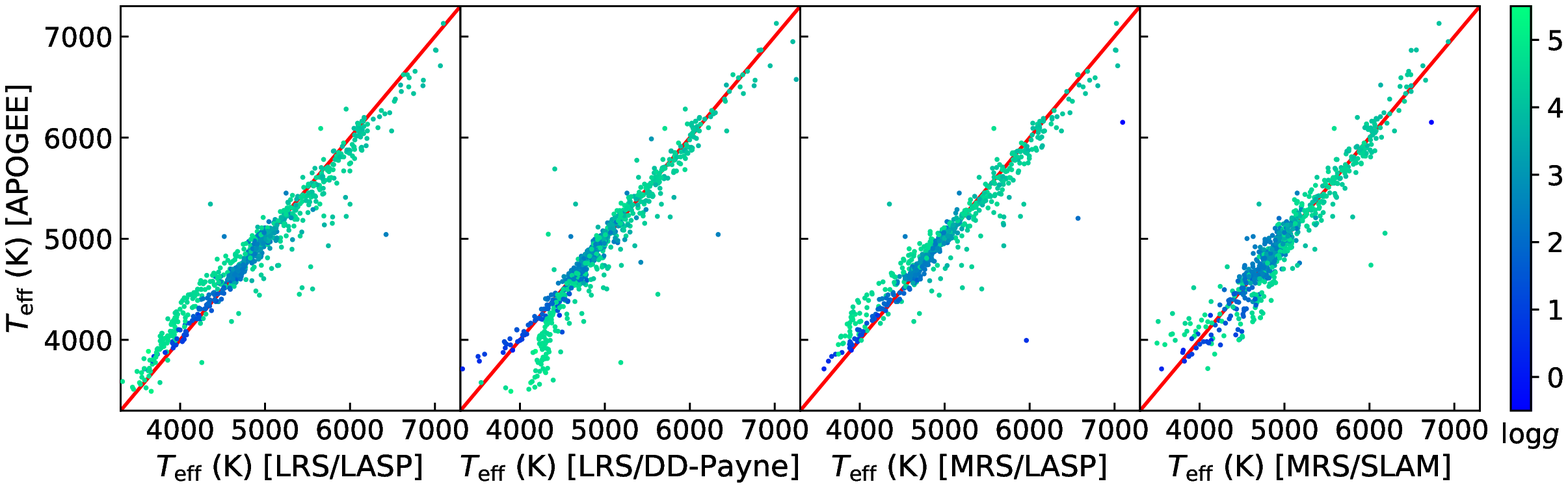}
   \includegraphics[width=0.98\textwidth]{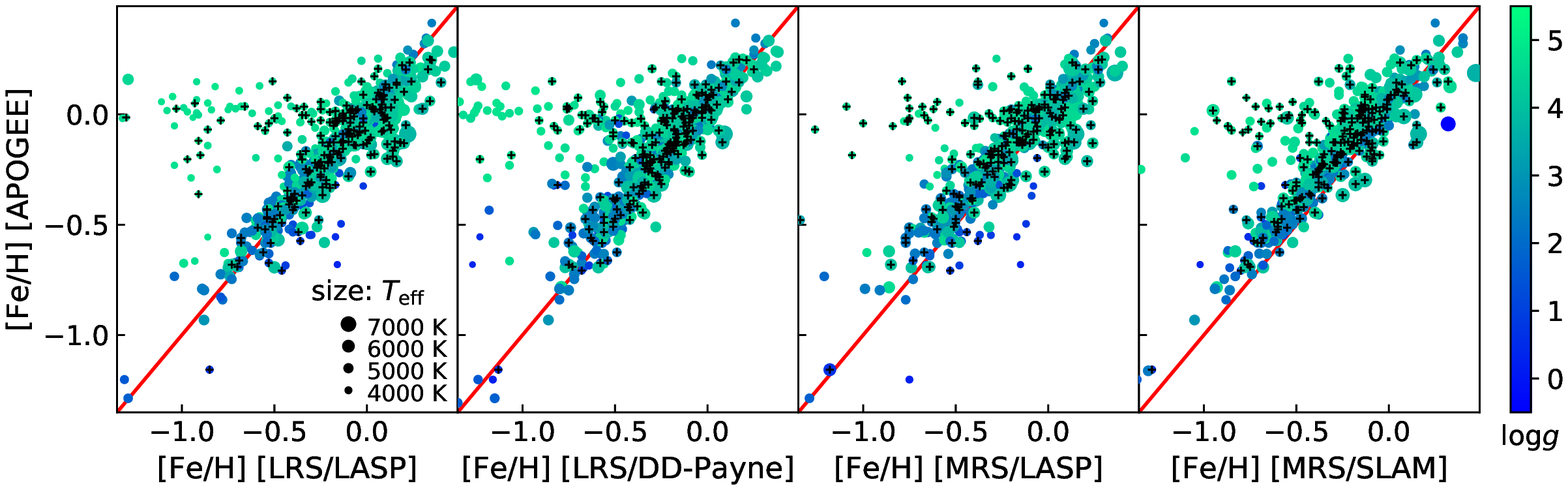}
   \caption{Top panel: Comparison of $T_\mathrm{eff}$ between LRS/LASP, LRS/DD-Payne, MRS/LASP, MRS/SLAM and APOGEE. The colorbar represents log$g$. Bottom Panel: Comparison of [Fe/H] between LRS/LASP, LRS/DD-Payne, MRS/LASP, MRS/SLAM and APOGEE. The colorbar represents log$g$. The size of symbols represents $T_{\rm eff}$. The black pluses are binary candidates (Section \ref{binary.sec}).}
   \label{comLA2.fig}
\end{figure*}

There are some outliers showing clear discrepancy of $T_\mathrm{eff}$ values.
For objects located in the range [4500, 6500]\,K, their $T_\mathrm{eff}$ values from different methods in this study are consistent with those from APOGEE (Figure \ref{comLA2.fig}).
For cooler dwarfs, the LASP returns lower temperature than APOGEE, while DD-Payne gives higher temperatures.
Some of these sources may be variable stars, since we preferred variable sources to construct our sample.
Inappropriate stellar templates may also result in inaccurate parameter measurements \citep{2020ApJS..251...15Z}.

Although most of the stars in common have consistent metallicities with each other,
we note that some objects show large discrepancy of [Fe/H] values (Figure \ref{comLA2.fig}).
Our methods derived much lower metallicity than those of APOGEE.
These sources are cool dwarfs ($T_{\rm eff} \lesssim$ 4000K; log$g$ $\gtrsim$ 4.5).
Some of these sources are probably variable stars or binaries, and clearly the parameter estimations of the latter are inaccurate.
On the other hand, it is difficult to determine accurate stellar parameters for very cool dwarfs.

\section{Mass Determination}
\label{mass.sec}

We determined an evolutionary mass by using two methods and a spectroscopic mass for our sample stars.
Since the LASP method was used to derive stellar parameters for both the LRS and MRS data, we preferred to use their parameter values, followed by the DD-Payne (for LRS data) and SLAM (for MRS data) results.

\subsection{evolutionary mass estimation}
We used the Modules for Experiments in Stellar Astrophysics \citep[{\sc MESA};][]{2011ApJS..192....3P,2013ApJS..208....4P, 2015ApJS..220...15P,2018ApJS..234...34P} (version 12115) to construct a grid of stellar models.
We calculated the initial chemical composition by using the solar chemical mixture [$(Z/X)_{\odot}$ = 0.0181] \citep{2009ARA&A..47..481A}.
The {\sc mesa} $\rho-T$ tables based on the 2005 update of the OPAL equation of state tables \citep{2002ApJ...576.1064R} was adopted and we used the OPAL opacities supplemented by the low-temperature opacities from \citet{2005ApJ...623..585F}. The {\sc mesa} Eddington photosphere was used for the set of boundary conditions for modelling the atmosphere. The mixing-length theory of convection was implemented and $\alpha_{\rm MLT}$ refers to the mixing-length parameter.
We also applied the {\sc mesa} predictive mixing scheme in our model for a smooth convective boundary.
We considered convective overshooting at the core, the H-burning shell, and the envelope.
The exponential scheme by \citet{2000A&A...360..952H} was applied.
The overshooting parameter is mass-dependent following a relation as $f_{\rm{ov}}$ = (0.13$M$ $-$ 0.098)$/$9.0 found by \citet{2010ApJ...718.1378M}. In addition, we adopted a fixed $f_{\rm{ov}}$ of 0.018 for models above $M = 2.0$\,M$_{\odot}$. The mass-loss rate on the red-giant branch with Reimers prescription was set as $\eta = 0.2$ as constrained by the seismic targets in old open clusters NGC\,6791 and NGC\,6819 \citep{2012MNRAS.419.2077M}.
Our models contain four independent inputs which are mass (M = 0.76 -- 2.2/0.02 $M_{\odot}$), initial helium fraction ($Y_{\rm init}$ = 0.24 -- 0.32/0.02), initial metallicity
([Fe/H]$_{\rm init}$ = $-$0.5 -- 0.5/0.1), and the mixing-length parameter
($\alpha_{\rm MLT}$ = 1.7 $-$ 2.3/0.2).
We used the maximum-likelihood estimation to fit to spectroscopic constraints to determine the stellar masses.

\begin{figure*}[!htbp]
   \center
   \includegraphics[width=0.98\textwidth]{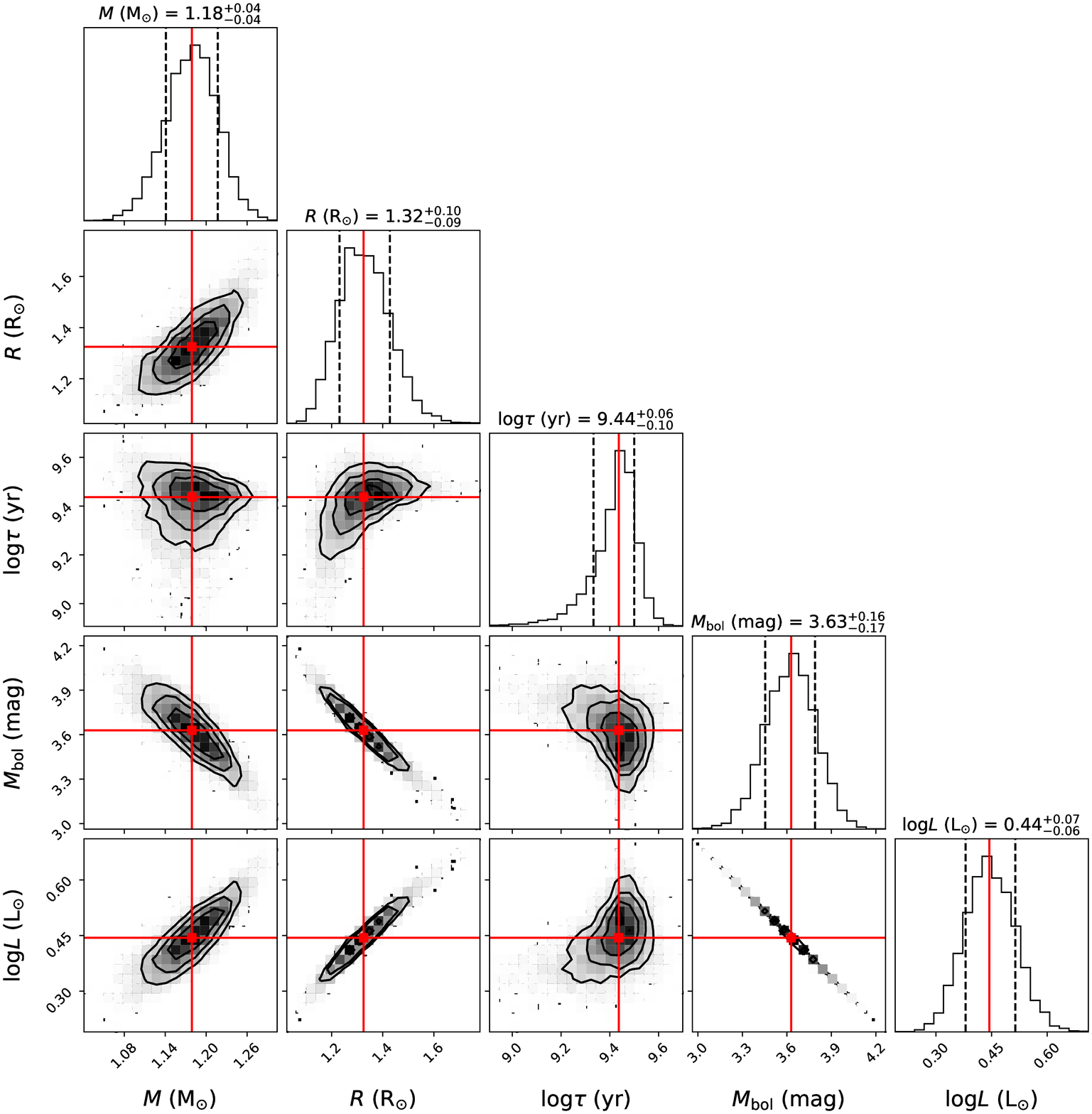}
   \caption{Corner plot showing the distribution of physical parameters of J034004.12+235200.0 as an example, derived from the {\it isochrones} code. The parameters are labeled as, mass ($M$, in M$_{\odot}$),  radius ($R$, in R$_{\odot}$), age (log$\tau$, in yr), bolometric magnitude ($M_{\rm bol}$, in mag), and bolometric luminosity (log$L$, in L$_{\odot}$).}
   \label{mistcorner.fig}
\end{figure*}

We also applied the ``{\it isochrones}" Python module \citep{2015ascl.soft03010M} to estimate stellar mass, which is an interpolation tool for the fitting of stellar models to photometric or spectroscopic parameters.
By using the trilinear interpolation in mass–age–[Fe/H] space for any given set of model grids, it is  able to predict physical or photometric properties provided by the models \citep{2015ApJ...809...25M}.
The input of the code includes the measured temperature, surface gravity, multi-band magnitudes ($G$, $G_{\rm BP}$, $G_{\rm RP}$, $J$, $H$, and $K_{\rm S}$), {\it Gaia} DR2 parallax \citep{2018AA...616A...1G} and reddening $E(B-V)$.
The $E(B-V)$ value is calculated with
$E(B-V) = 0.884 \times {\rm (Bayestar19)}$, with the latter\footnote{http://argonaut.skymaps.info/usage} from the Pan-STARRS DR1 (hereafter PS1) dust map \citep{2015ApJ...810...25G}.
An example of the fitting results is given in Figure \ref{mistcorner.fig}.

We remind that the evolutionary masses were calculated assumed no metal enrichment. There are about 1200/200 objects with [Fe/H] lower than $-$0.5/$-$1. Their masses may be under-estimated if there are significant $\alpha$-elements enrichment.

\subsection{Spectroscopic mass estimation}

The stellar mass can be estimated with the observed spectroscopic and photometric parameters.
First, We calculated an uncertainty-weighted average bolometric magnitude with Eq. (2) and (3), by using the multi-band magnitudes ($G$, $G_{\rm BP}$, $G_{\rm RP}$, $J$, $H$, and $K_{\rm S}$), the {\it Gaia} DR2 distance \citep{2018AJ....156...58B}, the extinction from PS1 dust map, and the bolometric corrections \citep{2019AA...632A.105C}.
For 2MASS magnitudes, we derived the attenuation by directly multiplying the extinction coefficients on PS1's website \footnote{http://argonaut.skymaps.info/usage} by the Bayestar19 value; for {\it Gaia} magnitudes, we calculated the $E(B-V)$ and derived the extinction by multiplying $E(B-V)$ by the extinction coefficients from \citet{2018MNRAS.479L.102C}.
The bolometric correction is derived from the PARSEC database \footnote{http://stev.oapd.inaf.it/YBC/}, with the input of $T_{\rm eff}$, log$g$, and [Fe/H] values.
Second, the bolometric luminosity was calculated with the averaged bolometric magnitude and the absolute luminosity and magnitude of the sun ($L_{\odot} =$ 3.83$\times$ 10$^{33}$ erg/s; $M_{\odot} =$ 4.74 mag).
Finally, we derived the stellar mass with the bolometric luminosity, effective temperature, and surface gravity following
\begin{equation}
M = \frac{L_{\rm bol}}{4\pi~G\sigma~T_{\rm eff}^{4}}g.  \\
\end{equation}

\begin{figure*}[!htbp]
   \center
   \includegraphics[width=0.49\textwidth]{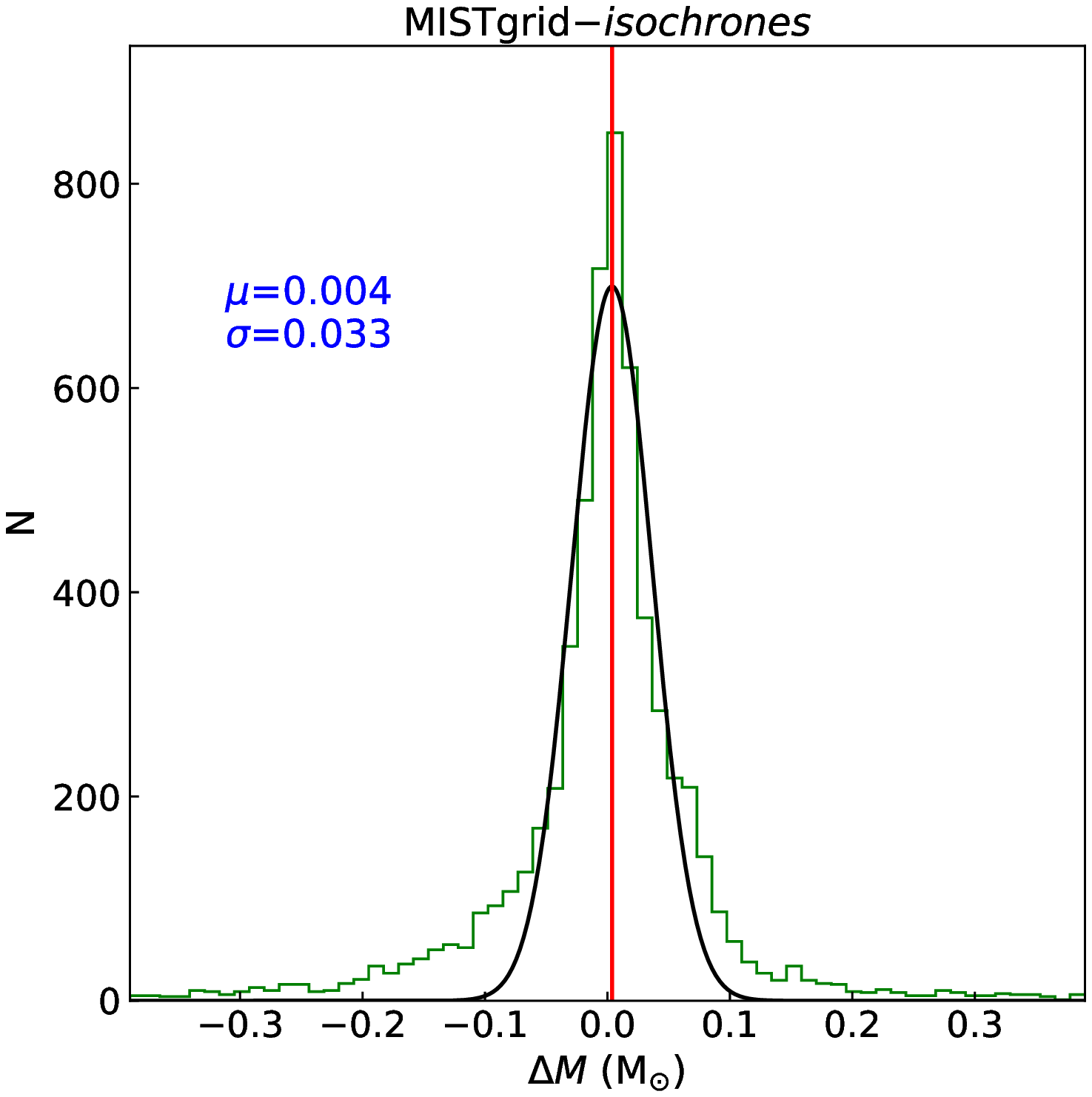}
   \includegraphics[width=0.49\textwidth]{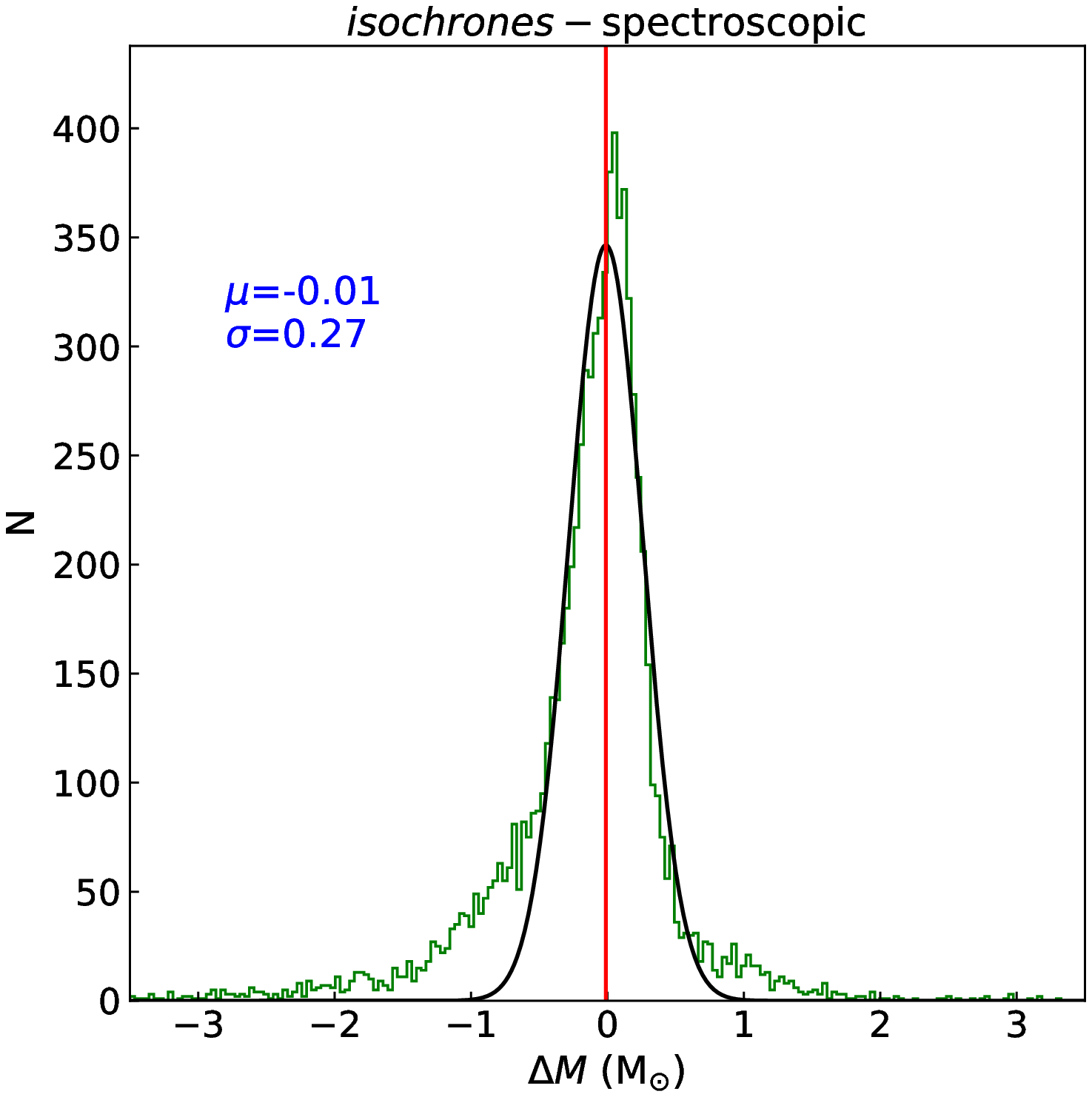}
   \caption{Left Panel: Comparison of the mass values from MISTgrid and {\it isochrones}. Right Panel: Comparison of the mass values from {\it isochrones} and spectroscopic estimation.}
   \label{masscom.fig}
\end{figure*}

\begin{figure*}[!htbp]
   \center
   \includegraphics[width=0.49\textwidth]{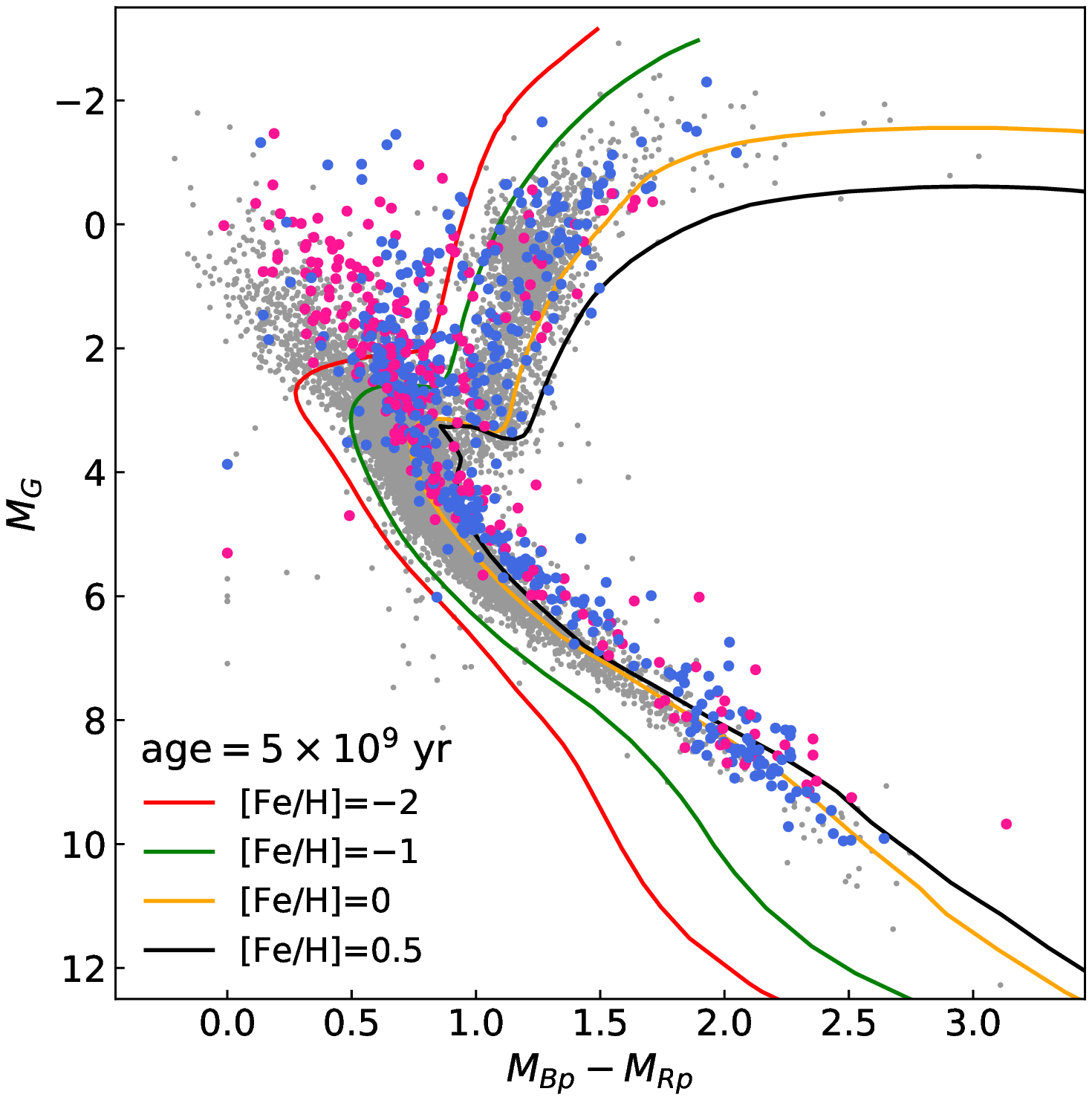}
   \includegraphics[width=0.49\textwidth]{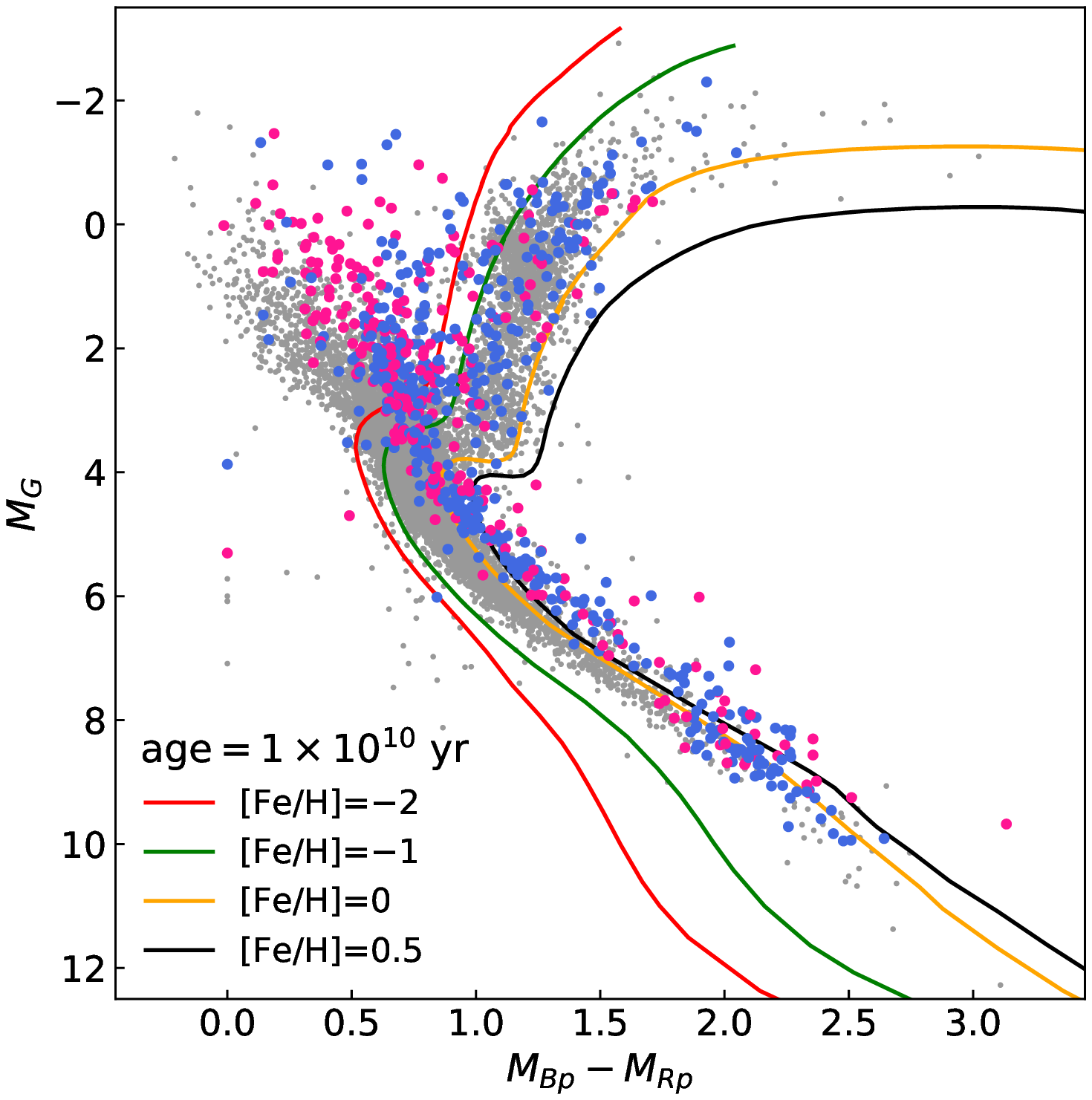}
   \caption{Left Panel: Color-magnitude diagram of the sample stars. The larger (red and blue) dots represent the stars with large mass discrepancy. The red dots are binary candidates from Section \ref{binary.sec}. Four isochrones are shown for comparison, which are at age of 5 billion year with different metallicities ([Fe/H] $=$  $-$2, $-$1, 0, 0.5). Right panel: Four isochrones are drawn, which are at age of 10 billion year with different metallicities ([Fe/H] $=$ $-$2, $-$1, 0, 0.5).}
   \label{mass.fig}
\end{figure*}

The comparison of mass estimation with MIST grid and {\it isochrones} shows good agreement (Figure \ref{masscom.fig}). However, some targets show higher spectroscopic mass than the evolutionary mass .
There are about 750 sources with $\lvert \Delta M \rvert / M_{iso}$ $\geq$ 1, and about 270 ones are in our binary catalog (Section \ref{binary.sec}).
In fact, most of these sources throughout the main sequence are probably unresolved binaries, since they are clearly brighter than the main-sequence stars with the same color (Figure \ref{mass.fig}).

\begin{table*}
\caption[]{Mass estimations of the sample stars. \label{mass.tab}}
 \begin{tabular}{ccccccc}
  \hline\noalign{\smallskip}
Name & $M_{\rm grid}$ & $M_{iso}$ &  Dis. & $E(B-V)$ &  $M_{\rm bol}$  &  $M_{\rm spec}$  \\
 &   (M$_{\odot}$) &   (M$_{\odot}$) & (kpc) &  &  (mag)  &  (M$_{\odot}$) \\
   \hline\noalign{\smallskip}
J034004.12+235200.0 & 1.02$_{-0.04}^{+0.04}$ & 1.18$\pm$0.04 & 279$_{-3}^{+4}$ & 0.19 & 3.83$\pm$0.04 & 0.94$\pm$0.14\\
J034007.72+241820.5 & --- & 0.99$\pm$0.1 & 903$_{-32}^{+34}$ & 0.21 & 4.04$\pm$0.01 & 0.7$\pm$0.03\\
J034008.18+241703.1 & 1.86$_{-0.24}^{+0.22}$ & 1.33$\pm$0.15 & 1319$_{-87}^{+100}$ & 0.21 & 0.11$\pm$0.05 & 2.87$\pm$0.13\\
J034012.25+234313.8 & --- & 1.13$\pm$0.34 & 2391$_{-163}^{+188}$ & 0.23 & 0.41$\pm$0.03 & 0.83$\pm$0.03\\
J034012.43+233803.1 & 1.24$_{-0.05}^{+0.04}$ & 1.21$\pm$0.03 & 471$_{-17}^{+19}$ & 0.19 & 2.57$\pm$0.08 & 2.5$\pm$0.67\\
J034020.87+234005.1 & 0.96$_{-0.04}^{+0.06}$ & 1.08$\pm$0.02 & 233$_{-4}^{+4}$ & 0.22 & 3.89$\pm$0.07 & 1.29$\pm$0.24\\
J034020.90+242455.5 & --- & 0.86$\pm$0.05 & 615$_{-13}^{+14}$ & 0.19 & 4.89$\pm$0.01 & 0.76$\pm$0.02\\
J034024.32+242932.1 & --- & 0.91$\pm$0.07 & 473$_{-15}^{+16}$ & 0.21 & 2.08$\pm$0.02 & 1.09$\pm$0.03\\
J034025.52+241017.1 & --- & 2.15$\pm$0.28 & 229$_{-2}^{+2}$ & 0.29 & 5.99$\pm$0.43 & 0.0$\pm$0.0\\
J034025.64+244209.5 & 1.24$_{-0.08}^{+0.08}$ & 1.25$\pm$0.08 & 1030$_{-32}^{+34}$ & 0.19 & 2.57$\pm$0.07 & 1.82$\pm$0.42\\
J034025.96+232013.5 & --- & 0.96$\pm$0.03 & 1731$_{-375}^{+569}$ & 0.19 & 1.63$\pm$0.28 & 7.75$\pm$5.95\\
J034026.48+235823.3 & --- & 1.08$\pm$0.21 & 4197$_{-464}^{+574}$ & 0.28 & -0.47$\pm$0.04 & 1.87$\pm$0.06\\
J034029.22+234840.1 & 0.92$_{-0.06}^{+0.06}$ & 0.94$\pm$0.01 & 237$_{-32}^{+43}$ & 0.22 & 3.51$\pm$0.07 & 5.13$\pm$0.72\\
J034029.59+233303.9 & --- & 0.59$\pm$0.06 & 161$_{-1}^{+1}$ & 0.21 & 7.01$\pm$0.24 & 0.52$\pm$0.07\\
J034030.72+242914.2 & --- & 0.9$\pm$0.01 & 138$_{-1}^{+1}$ & 0.03 & 5.68$\pm$0.08 & 1.29$\pm$0.15\\
J034031.01+242141.0 & --- & 0.38$\pm$0.02 & 69$_{-0}^{+0}$ & 0.2 & 7.68$\pm$0.15 & 0.61$\pm$0.04\\
J034031.67+234521.9 & 0.92$_{-0.06}^{+0.06}$ & 0.98$\pm$0.05 & 710$_{-19}^{+20}$ & 0.22 & 4.49$\pm$0.03 & 0.84$\pm$0.06\\
J034031.88+243419.1 & 1.3$_{-0.32}^{+0.36}$ & 1.35$\pm$0.38 & 2101$_{-138}^{+159}$ & 0.21 & 1.46$\pm$0.04 & 1.57$\pm$0.08\\
J034034.36+234057.3 & 0.98$_{-0.02}^{+0.02}$ & 1.05$\pm$0.01 & 134$_{-1}^{+1}$ & 0.12 & 4.94$\pm$0.08 & 1.15$\pm$0.2\\
J034034.76+232540.2 & 1.56$_{-0.14}^{+0.14}$ & 1.74$\pm$0.16 & 1534$_{-65}^{+71}$ & 0.18 & 1.89$\pm$0.11 & 1.57$\pm$1.44\\
  \noalign{\smallskip}\hline
\end{tabular}
\smallskip\\
\tablecomments{\textwidth}{The columns are: (1) Name; (2) $M_{\rm grid}$: mass estimation from the MIST grids; (3) $M_{\rm iso}$: mass estimation using the ``isochrones" code; (4) Dis.: distance from {\it Gaia} DR2; (5) $E(B-V)$: reddening from PS1 dust map, calculated as 0.84$\times$Bayesian19; (6) $M_{\rm bol}$: weighted average value of bolometric magnitude; (7) $M_{\rm spec}$: spectroscopic mass estimation.\\
This table is available in its entirety in machine-readable and Virtual Observatory (VO) forms in the online journal. A portion is shown here for guidance regarding its form and content.}
\end{table*}

\section{Binary sample}
\label{binary.sec}

We present a binary sample basing on light curve analysis, radial velocity fitting, the binarity parameter calculated using MRS data (Section \ref{slam.sec}), and the spatially resolved binary catalog from Gaia EDR3 \citep{2021MNRAS.tmp..394E}.
In this study, one star is thought to be a double-lined spectroscopic binary candidate if the binarity parameters of three more spectra (with SNR above 10) are larger than 0.9 (Figure \ref{binarity.fig}).

\subsection{light curve analysis}

We first cross-matched our catalog with the $K$2 data, and found more than 3000 stars have light curves, which can be used to detect period signals.
The moving average method was used to smooth the light curve and remove the long-term trend.
We applied the Lomb-Scargle method \citep{1976ApSS..39..447L} to determine the period and classified binaries by analyzing the folded light curves \citep[see more details in][]{2020ApJS..249...31Y}.
A brief description was presented as follows.

We used a two-step grid searching method \citep{2015ApJ...812...18V} to determine the optimized period.
It firstly searches in a broad grid for a series of period candidates and then zooms in a narrow grid to find the real peak. The obtained period is regarded as significant only when it is higher than the false alarm probability. The light curve folded with the significant period was analyzed by investigating the characteristics. The light curve templates of variable stars were taken from previous catalogs \citep[e.g.][]{2017ARep...61...80S,2014A&A...566A..43K}.
The characters of the templates include light curve period, skewness of the magnitude distribution, median magnitude, standard deviation of the magnitude, the ratio of magnitudes brighter or fainter than the average, the ratio between the Fourier components a$_{2}$ and a$_{4}$, 10$\%$ and 90$\%$ percentile of slopes of a phase-folded light curve.
They were concluded as identification parameters that trigger the classification through machine learning method and visual inspection \citep{2006MNRAS.368.1311P,2016A&A...587A..18K,2020MNRAS.491...13J,YangLTD}.

\begin{figure*}[!htbp]
   \center
   \includegraphics[width=0.49\textwidth]{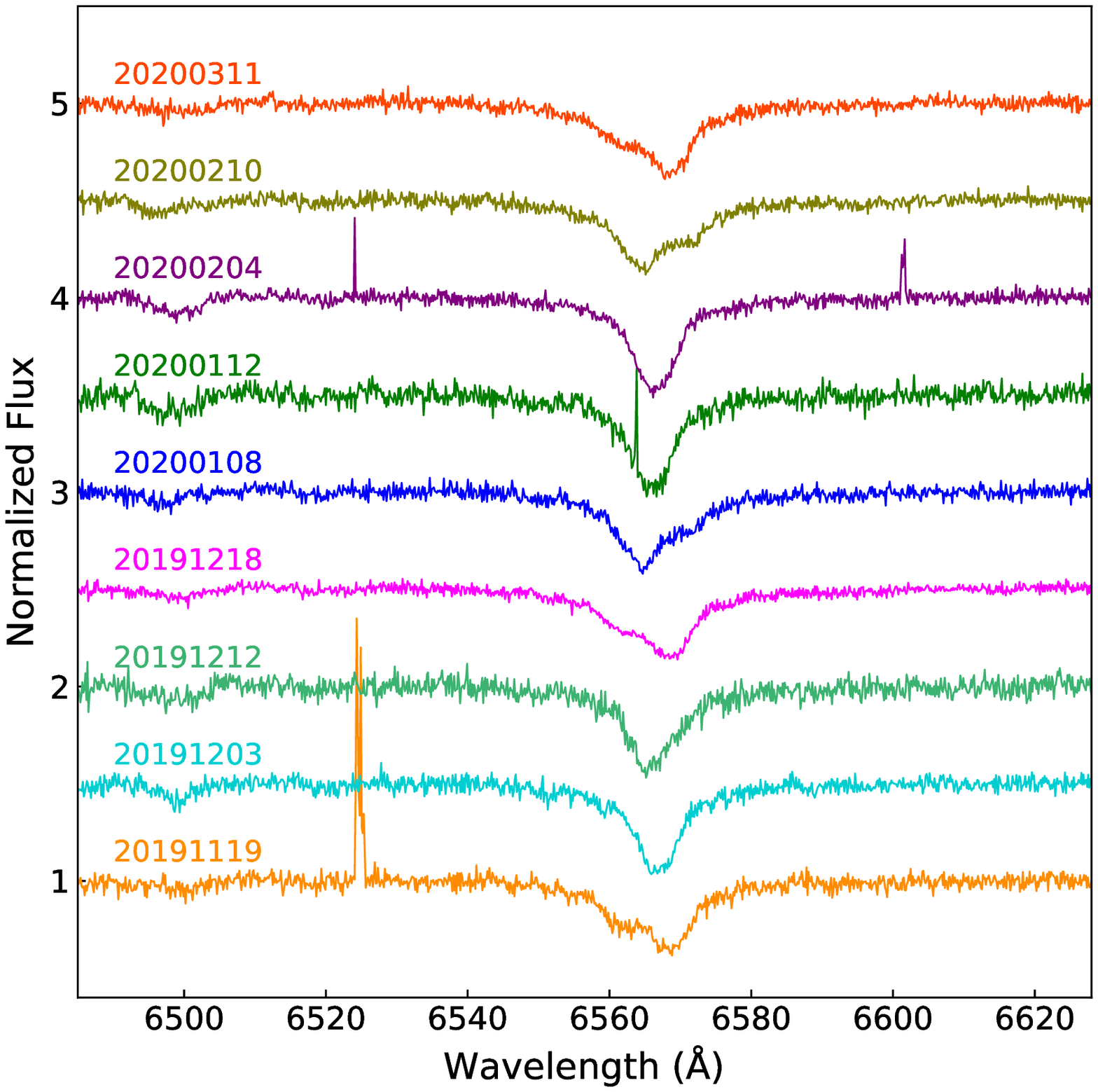}
   \includegraphics[width=0.49\textwidth]{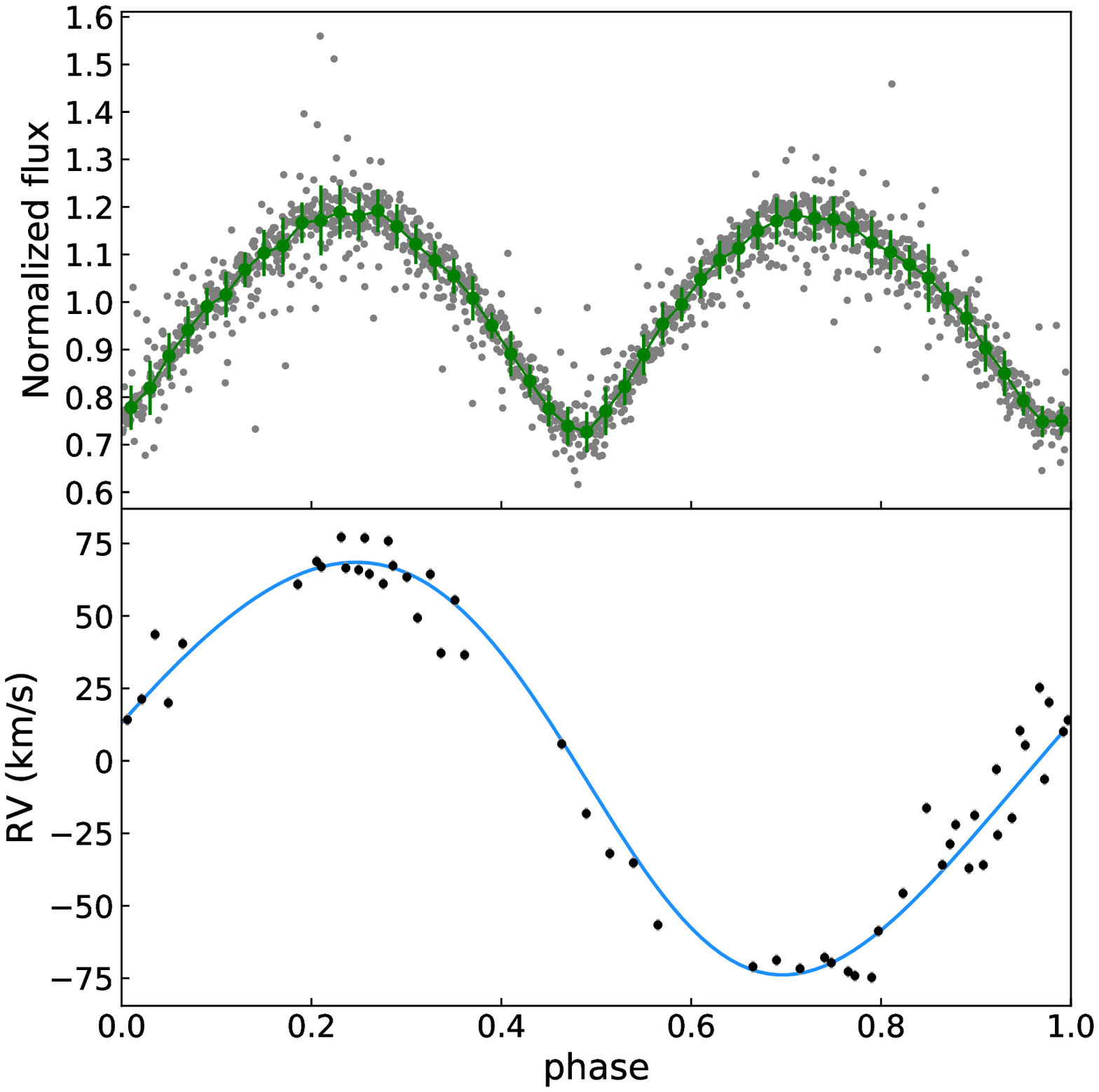}
   \caption{Left Panel: Part of MRS observations of J065001.65+222127.7 as an example. The binary feature can be clearly seen from the $H{\alpha}$ line profiles and motion. Right Panel: The folded light curve using $K$2 data and RV curve fitted with {\it The Joker}.}
   \label{binary.fig}
\end{figure*}

We also cross-matched our objects with the variable catalogs of ASAS-SN \footnote{https://asas-sn.osu.edu/variables}, Catalina, ZTF \citep{2020ApJS..249...18C}, and WISE \citep{2018ApJS..237...28C}.
Table \ref{binary.tab} shows the binaries with different types (i.e., EA, EB, and EW).
Figure \ref{binary.fig} shows a binary example of the EW type.

\subsection{radial velocity fitting}

\begin{figure*}[htbp!]
\center
\includegraphics[width=1\textwidth]{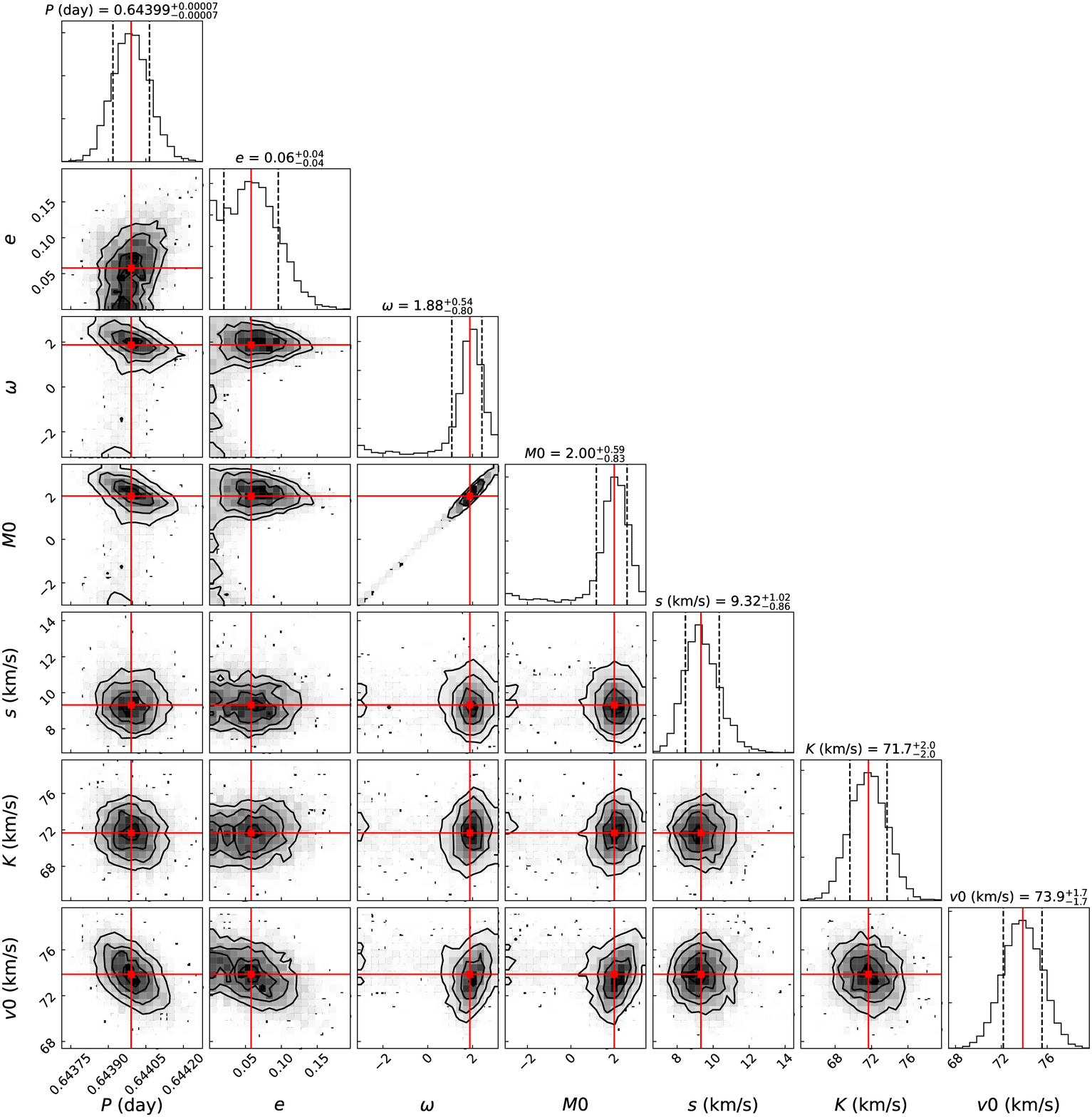}
\caption[]{Corner plot of the RV fitting results of J065001.65+222127.7 as an example, showing distribution of  orbital parameters derived from {\it The Joker}. The parameters are labeled as, orbital period ($P$, in days),  eccentricity of the system ($e$), argument of pericenter ($\omega$, in radians), mean anomaly at reference time ($M$0, in radians), extra ``jitter" added in quadrature to each visit-velocity error ($s$, in km/s), RV semi-amplitude of the star ($K$, in km/s), and the center of mass velocity ($\nu$0, in km/s).}
\label{jokermcmc.fig}
\end{figure*}

With the radial velocity data from the LAMOST TD survey, we performed a Keplerian fit using the custom Markov chain Monte Carlo sampler {\it The Joker} \citep{2017ApJ...837...20P} for the objects with more than seven exposures.
{\it The Joker} works well with non-uniform data and allows to identify circular or eccentric orbits.
We used the RVs of single-exposure spectra to do the fitting.
Four sets of data were used: the LRS RV from LASP, the MRS RV from LASP, the MRS RV from SLAM, and a joint LRS and MRS RV from LASP.
Bad fittings were removed with visual examination.
The fitting with MRS RV data was preferred, followed by the fitting with the joint data and the LRS RV data.
The derived orbital parameters include period $P$, eccentricity $e$, semi-amplitude $K$, argument of the periastron $\omega$, mean anomaly at the first exposure, and systematic RV $\nu$0.
An example of the fitting results is given in Figure \ref{jokermcmc.fig}.
The results are shown in Table \ref{binary.tab}.
For double-lined spectroscopic binaries, we only used the set of RV with larger semi-amplitude ($K$) to do the fitting.

In addition, for single-line binaries, which are not classified as binaries by the binarity parameter, we calculated the binary mass function $f(M)$ using the posterior samples from our RV modeling as follows,
\begin{equation}
    f(M) = \frac{P \, K_{1}^{3} \, (1-e^2)^{3/2}}{2\pi G} = \frac{M_{2} \, \textrm{sin}^3 i} {(1+q)^{2}},
\end{equation}
\noindent
where $K$1 is the semi-amplitude of the primary (i.e., the visible star), $M_{2}$ is the mass of the secondary, $q = M_{1}/M_{2}$ is the mass ratio, and $i$ is the system inclination.
Combined with the mass estimate of the primary (Section \ref{mass.sec}), we estimated a minimum mass of the secondary ($M$2) with an inclination angle of $i\ =\ $ 90$^{\circ}$.

\subsection{Spatially resolved binary}
By using the {\it Gaia} EDR3 database, \citet{2021MNRAS.tmp..394E} searched pairs of stars and estimated the probability that a pair is a chance alignment. They constructed a catalog of 1.2 million high-confidence, spatially resolved wide binaries.
We cross-matched our sample and their catalog, and found 379 common sources. Among these objects, 306 ones were classified as main sequence -- main sequence (MSMS) binaries and three ones were distinguished as white dwarf -- main sequence (WDMS) binaries.

To sum up, Table \ref{binary.tab} lists 2366 binary candidates, including 148 ones from light curve analysis, 878 ones from RV fitting, 1534 ones from binarity parameter, and 379 ones from the spatially resolved catalog of {\it Gaia} EDR3.

\begin{sidewaystable*}[h]
\setlength{\tabcolsep}{1.2pt}
\small
\center
\vspace{12cm}
\caption[]{Binary candidates of the four $K$2 plates. \label{binary.tab}}
 \begin{tabular}{cccccccccccccc}
  \hline\noalign{\smallskip}
 &  \multicolumn{8}{c}{{\it The Joker}} &   \multicolumn{2}{c}{light curve} & binarity $\geq$0.9 & \multicolumn{2}{c}{Astrometry} \\
\cmidrule(lr){2-9} \cmidrule(lr){10-11} \cmidrule(lr){13-14}
Name      &         $P$  &   $e$       & $\omega$&  $M_{\rm 0}$ &    $K$    & $\nu_{\rm 0}$  &    $f$($M$) & $M$2$_{\rm min}$ & Type/Survey  &  $P$   & & Class & Sep.   \\
 &   (day) &   &  &   &  (km/s)  &  (km/s)  & M$_{\odot}$ &  M$_{\odot}$ & & (day) &  & & (AU)\\
  \hline\noalign{\smallskip}
J034012.43+233803.1 & --- & --- & --- & --- & --- & --- & --- & --- & EA/AAVSO & 17.3679 & --- & --- & ---\\
J034025.64+244209.5 & 1.4740$_{-0.0005}^{+0.0004}$ & 0.352$_{-0.063}^{+0.083}$ & 1.88$_{-0.15}^{+0.15}$ & -1.34$_{-0.27}^{+0.24}$ & 39.7$_{-1.8}^{+2.5}$ & -16.5$_{-2.1}^{+2.6}$ & 0.0078$_{-0.0008}^{+0.0011}$ & 0.26$_{-0.02}^{+0.02}$ & --- & --- & --- & --- & ---\\
J034031.01+242141.0 & --- & --- & --- & --- & --- & --- & --- & --- & --- & --- & 1.00 & --- & ---\\
J034051.81+232834.4 & 20.0490$_{-0.0411}^{+0.0508}$ & 0.514$_{-0.025}^{+0.028}$ & -0.10$_{-0.06}^{+0.05}$ & 0.99$_{-0.08}^{+0.07}$ & 19.8$_{-0.7}^{+0.7}$ & 4.8$_{-0.4}^{+0.4}$ & 0.0102$_{-0.0010}^{+0.0011}$ & 0.29$_{-0.01}^{+0.01}$ & --- & --- & --- & --- & ---\\
J034100.15+241735.0 & --- & --- & --- & --- & --- & --- & --- & --- & --- & --- & --- & MSMS & 694\\
J034108.07+231255.5 & 0.3349$_{-0.0002}^{+0.0700}$ & 0.370$_{-0.206}^{+0.230}$ & 4.10$_{-5.97}^{+0.68}$ & 2.85$_{-4.76}^{+1.19}$ & 5.9$_{-1.5}^{+1.6}$ & 18.4$_{-0.7}^{+0.7}$ & 0.0001$_{-0.0001}^{+0.0001}$ & 0.01$_{0.00}^{+0.00}$ & --- & --- & --- & --- & ---\\
J034115.98+225250.0 & --- & --- & --- & --- & --- & --- & --- & --- & --- & --- & 0.96 & --- & ---\\
J034122.94+233730.6 & 2.2546$_{-0.0792}^{+0.2923}$ & 0.301$_{-0.166}^{+0.129}$ & 1.69$_{-0.19}^{+0.59}$ & 0.83$_{-0.41}^{+1.14}$ & 44.3$_{-9.1}^{+3.7}$ & -6.2$_{-8.1}^{+2.1}$ & 0.0150$_{-0.0052}^{+0.0109}$ & 0.32$_{-0.06}^{+0.09}$ & --- & --- & --- & --- & ---\\
J034125.62+240919.9 & 7.6155$_{-0.0224}^{+6.7914}$ & 0.218$_{-0.131}^{+0.646}$ & 3.83$_{-6.09}^{+1.15}$ & 0.94$_{-4.04}^{+2.25}$ & 18.7$_{-1.5}^{+2.0}$ & -31.9$_{-3.9}^{+1.6}$ & 0.0047$_{-0.0035}^{+0.0015}$ & 0.20$_{-0.09}^{+0.04}$ & --- & --- & --- & --- & ---\\
J034134.72+230542.7 & 1.2253$_{-0.0005}^{+0.0187}$ & 0.481$_{-0.050}^{+0.047}$ & 3.85$_{-0.17}^{+0.15}$ & -2.15$_{-0.26}^{+5.93}$ & 9.2$_{-0.6}^{+1.9}$ & -40.1$_{-0.4}^{+0.8}$ & 0.0001$_{-0.0001}^{+0.0001}$ & 0.04$_{-0.01}^{+0.01}$ & --- & --- & --- & --- & ---\\
J034137.08+230049.9 & --- & --- & --- & --- & --- & --- & --- & --- & --- & --- & --- & MS?? & 643\\
J034141.71+241910.1 & 5.2130$_{-0.7949}^{+0.0109}$ & 0.266$_{-0.082}^{+0.200}$ & 5.81$_{-0.49}^{+0.28}$ & 0.99$_{-0.75}^{+0.40}$ & 21.4$_{-2.0}^{+3.0}$ & 14.6$_{-1.4}^{+1.5}$ & 0.0047$_{-0.0016}^{+0.0019}$ & 0.19$_{-0.03}^{+0.03}$ & --- & --- & --- & --- & ---\\
J034144.50+232159.4 & 4.6605$_{-0.0080}^{+0.0064}$ & 0.231$_{-0.059}^{+0.070}$ & -2.54$_{-0.31}^{+0.26}$ & -3.79$_{-0.40}^{+0.40}$ & 20.7$_{-1.0}^{+1.0}$ & -28.3$_{-0.9}^{+1.0}$ & --- & --- & --- & --- & 1.00 & --- & ---\\
J034145.06+231235.2 & 0.4943$_{-0.0001}^{+0.0001}$ & 0.022$_{-0.017}^{+0.037}$ & 0.67$_{-1.67}^{+1.90}$ & 0.34$_{-1.67}^{+1.79}$ & 60.1$_{-1.2}^{+1.2}$ & -9.6$_{-1.9}^{+2.0}$ & --- & --- & EB/ASASSN & 0.4942 & 1.00 & --- & ---\\
J034148.27+224912.0 & --- & --- & --- & --- & --- & --- & --- & --- & --- & --- & 1.00 & --- & ---\\
J034154.04+224222.6 & --- & --- & --- & --- & --- & --- & --- & --- & --- & --- & --- & MSMS & 10322\\
J034205.65+233515.8 & 27.7518$_{-0.1930}^{+0.1695}$ & 0.360$_{-0.038}^{+0.040}$ & 2.82$_{-0.17}^{+0.16}$ & -0.20$_{-0.13}^{+0.12}$ & 10.9$_{-0.3}^{+0.4}$ & -55.3$_{-0.4}^{+0.4}$ & 0.0030$_{-0.0003}^{+0.0004}$ & 0.18$_{-0.01}^{+0.01}$ & --- & --- & --- & --- & ---\\
J034209.14+233004.9 & 12.4971$_{-0.0076}^{+0.0075}$ & 0.028$_{-0.016}^{+0.016}$ & 1.45$_{-0.48}^{+0.58}$ & -0.09$_{-0.48}^{+0.59}$ & 52.4$_{-1.1}^{+1.1}$ & -12.3$_{-0.6}^{+0.6}$ & --- & --- & --- & --- & 1.00 & --- & ---\\
J034210.91+240508.6 & 0.2583$_{-0.0142}^{+0.0536}$ & 0.316$_{-0.251}^{+0.353}$ & 1.06$_{-0.94}^{+3.07}$ & 0.17$_{-1.01}^{+2.60}$ & 8.9$_{-6.0}^{+7.0}$ & 7.7$_{-2.5}^{+4.1}$ & 0.0001$_{-0.0001}^{+0.0001}$ & 0.01$_{0.00}^{+0.00}$ & --- & --- & --- & --- & ---\\
J034220.18+242806.1 & --- & --- & --- & --- & --- & --- & --- & --- & --- & --- & 1.00 & MSMS & 14373\\
  \noalign{\smallskip}\hline
\end{tabular}
\smallskip\\
{This table is available in its entirety in machine-readable and Virtual Observatory (VO) forms in the online journal. A portion is shown here for guidance regarding its form and content.}
\end{sidewaystable*}

\section{Discussion and Summary}
\label{summary.sec}

With one year of LAMOST observations, our project acquired more than 767,000 low- and 478,000 med-resolution spectra, corresponding to a total exposure time of $\approx$46.7 and $\approx$49.1 hours, respectively.
More than 70\%/50\% of low-resolution/med-resolution spectra have SNR above 10.

We determined stellar parameters (e.g., $T_{\rm eff}$, log$g$, [Fe/H]) and RV by using different methods (i.e., LASP, DD-Payne, and SLAM), and derived SNR-weighted average values of these parameters for our targets.
Generally, these parameters determined from different methods show good agreement, especially for late F-, G-, and early K-type stars.
The LRS and MRS results show a discrepancy of the RV measurements ($\approx$5.5 km/s).
The comparison of stellar parameters with APOGEE DR16 show good agreement, but the RV values from LRS data show a large discrepancy ($\approx$6.5 km/s) with those of APOGEE.
We used the {\it Gaia} DR2 RV data to calculate a median RVZP for each spectrograph exposure by exposure, and the RVZP-corrected RVs agree very well with those of APOGEE DR16.
We derived stellar masses by using different methods (i.e., MIST grids, {\it isochrones} code, and spectroscopic estimation), with the help of stellar parameters, multi-band magnitudes, distances and extinction values.

Basing on light curve analysis, radial velocity fitting, the binarity parameter, and the spatially resolved binary catalog from Gaia EDR3, we presented a binary catalog including about 2700 candidates.
We should remind that we derived stellar parameters and masses assuming the target is a single star, which
means for the binary candidates, these parameter values may be unreliable.

Our spectroscopic survey has gained multiple visits (up to 86 LRS visits and 54 MRS visits) for about 10,000 stars,
which can effectively leverage sciences in various research fields, such as

(1) Binary system. The monitoring of RV variation can reveal a large sample of binaries, especially those double-lined spectroscopic binaries. The time-series variations of RV, together with the light curves from photometric surveys, can help to determine the orbital properties, including the period, eccentricity, inclination angle, etc.
The statistical properties of the binaries (e.g., period, eccentricity, and metallicity) can provide critical clues on the formation and evolution of the binary systems. In addition, LAMOST has showed the ability to discover fantastic binaries, such as compact binaries including a neutron star or black hole \citep{2019Natur.575..618L}. Those binaries are greatly helpful in understanding the late evolution of massive stars, such as the formation of type Ia supernovae.
An analysis of the binaries in the four plates, including stellar parameter estimation for individual component, will be presented in a future work (Kovalev et al. 2021, in prep).

(2) Stellar activity. Many studies have focused on the evolution of stellar photospheric activity with spots or flares, by using the photometric TD survey data. In contrast, due to the lack of long-term spectroscopic observation, the evolution of chromospheric activity was studied for only a few stars. LAMOST TD survey provides a great opportunity to investigate stellar chromospheric activity over a large sample of stars with different spectral types, the variation of chromospheric activity due to rotational modulation of a single star or orbital modulation of a binary system, and the long-term evolution of chromospheric activity. All of these are quite helpful to understand the stellar magnetic activity and dynamo mechanism. An analysis of the stellar chromospheric activities using Ca II H\&K and Balmer lines is under way (Han et al. 2021, in prep).

(3) Stellar pulsation. Asteroseismology is a unique technique to study the internal physics of pulsating stars. Precise atmospheric parameters from LAMOST multiple spectral observations can help to constrain the parameter space in seismic searches for an optimal model. Periodic variation of atmospheric parameters and RV due to pulsation provide good opportunity to probe the dynamical processes of pulsation.

The LAMOST TD data can also be used in many other fields, such as studying the chemical abundance of special stars (e.g., metal-poor star, Lithium-rich star), investigating the spatial structure of the Galaxy together with the {\it Gaia} astrometric data, etc.

All the spectra used in this study are now available in the LAMOST DR8. The observations of the four $K$2 plates will be carried on but with reduced visiting frequency. At the same time, a similar TD survey of another four $K$2 plates is being carried out.

\normalem
\begin{acknowledgements}

We thank the anonymous referee for helpful comments and suggestions that have  improved the paper. 
Guoshoujing Telescope (the Large Sky Area Multi-Object Fiber Spectroscopic Telescope LAMOST) is a National Major Scientific Project built by the Chinese Academy of Sciences. Funding for the project has been provided by the National Development and Reform Commission. LAMOST is operated and managed by the National Astronomical Observatories, Chinese Academy of Sciences.
This work presents results from the European Space Agency (ESA) space mission Gaia. Gaia data are being processed by the Gaia Data Processing and Analysis Consortium (DPAC). Funding for the DPAC is provided by national institutions, in particular the institutions participating in the Gaia MultiLateral Agreement (MLA). The Gaia mission website is https://www.cosmos.esa.int/gaia. The Gaia archive website is https://archives.esac.esa.int/gaia.
In this work, we have made use of SDSS-IV APOGEE DR16 data. Funding for the Sloan Digital Sky Survey IV has been provided by the Alfred P. Sloan Foundation, the U.S. Department of Energy Office of Science, and the Participating Institutions. SDSS-IV acknowledges support and resources from the Center for High-Performance Computing at the University of Utah. The SDSS website is www.sdss.org. SDSS is managed by the Astrophysical Research Consortium for the Participating Institutions of the SDSS Collaboration which are listed at www.sdss.org/collaboration/affiliations/.
This publication makes use of data products from the Pan-STARRS1 Surveys (PS1) and the PS1 public science archive,
and the Two Micron All Sky Survey.
We acknowledge use of the VizieR catalogue access tool, operated at CDS, Strasbourg, France, and of Astropy, a community-developed core Python package for Astronomy (Astropy Collaboration, 2013).
This work was supported by the National Science Foundation of China (NSFC) under grant numbers 11988101, 11933004, and 12003050.
It was also supported by the National Key Research and Development Program of China (NKRDPC) under grant numbers 2019YFA0405000, 2019YFA0405504, and 2016YFA0400804, and the B-type Strategic Priority Program of the Chinese Academy of Sciences under grant number XDB41000000.
S. W. and H.-L. Y. acknowledges support from the Youth Innovation Promotion Association of the CAS (id. 2019057 and 2020060, respectively).
T. L. acknowledges the funding from the European Research Council (ERC) under the European Union Horizon 2020 research and innovation programme (CartographY GA. 804752).

\end{acknowledgements}

\bibliographystyle{raa}
\bibliography{bibtex}

\appendix
\section{Stellar Parameter catalogs}

We present the weight-averaged stellar parameters from different methods here. Table \ref{par_houlow.tab}, \ref{par_houmed.tab}, \ref{par_xianglow.tab}, and \ref{par_zhangmed.tab} are from LASP estimation with LRS data, LASP estimation with MRS data, DD-Payne estimation with LRS data, and SLAM estimation with MRS data, respectively.
Details of the different methods can be found in Section \ref{par.sec}.

\begin{table*}
\caption[]{Stellar parameters and RV from LASP estimation with LRS data. \label{par_houlow.tab}}
\setlength{\tabcolsep}{2pt}
\footnotesize
 \begin{tabular}{ccccccccc}
  \hline\noalign{\smallskip}
Name & Field & R.A. &  Dec. & $T_{\rm eff}$ &  log$g$  &  [Fe/H]  &  RV  &  corrected RV\\
 &  & (deg) &  (deg) & (K) &   &    &  (km/s)  &  (km/s) \\
   \hline\noalign{\smallskip}
J034004.12+235200.0 & TD035052N235741K01 & 55.0172 & 23.86668 & 6492$\pm$24 & $4.25\pm$0.02 & $-0.3\pm$0.02 & $-23.29\pm$4.88 & $-12.49\pm$3.38\\
J034007.72+241820.5 & TD035052N235741K01 & 55.03221 & 24.3057 & 5827$\pm$77 & $4.02\pm$0.13 & $-0.35\pm$0.18 & $-2.5\pm$2.24 & $4.86\pm$2.26\\
J034008.18+241703.1 & TD035052N235741K01 & 55.03411 & 24.2842 & 4577$\pm$14 & $2.64\pm$0.04 & $0.14\pm$0.03 & $-44.89\pm$1.63 & $-37.84\pm$1.48\\
J034012.25+234313.8 & TD035052N235741K01 & 55.05106 & 23.72051 & 4867$\pm$33 & $2.33\pm$0.09 & $-0.49\pm$0.03 & $-34.68\pm$2.25 & $-24.65\pm$1.54\\
J034012.43+233803.1 & TD035052N235741K01 & 55.05184 & 23.63421 & 6295$\pm$21 & $4.12\pm$0.03 & $-0.22\pm$0.02 & $1.19\pm$7.95 & $11.31\pm$6.36\\
J034020.87+234005.1 & TD035052N235741K01 & 55.087 & 23.66809 & 5909$\pm$23 & $4.25\pm$0.02 & $0.05\pm$0.02 & $40.02\pm$8.11 & $50.39\pm$9.7\\
J034020.90+242455.5 & TD035052N235741K01 & 55.08714 & 24.41547 & 5840$\pm$133 & $4.4\pm$0.18 & $-0.65\pm$0.07 & $-15.58\pm$3.19 & $-8.88\pm$3.15\\
J034024.32+242932.1 & TD035052N235741K01 & 55.10136 & 24.49227 & 4875$\pm$27 & $3.12\pm$0.07 & $-0.55\pm$0.05 & $-102.41\pm$2.3 & $-95.37\pm$1.9\\
J034025.64+244209.5 & TD035052N235741K01 & 55.10687 & 24.70268 & 6367$\pm$86 & $4.0\pm$0.1 & $-0.19\pm$0.08 & $-21.02\pm$22.58 & $-14.18\pm$22.82\\
J034025.96+232013.5 & TD035052N235741K01 & 55.10817 & 23.33711 & 6001$\pm$43 & $4.15\pm$0.06 & $-0.34\pm$0.04 & $50.43\pm$2.08 & $60.99\pm$2.35\\
J034026.48+235823.3 & TD035052N235741K01 & 55.11036 & 23.97316 & 4538$\pm$42 & $2.21\pm$0.12 & $-0.45\pm$0.05 & $-48.98\pm$2.91 & $-39.46\pm$2.85\\
J034029.22+234840.1 & TD035052N235741K01 & 55.1218 & 23.81117 & 5556$\pm$26 & $4.59\pm$0.02 & $0.03\pm$0.02 & $51.51\pm$2.56 & $62.3\pm$0.88\\
J034029.59+233303.9 & TD035052N235741K01 & 55.12332 & 23.55111 & 3999$\pm$26 & $4.43\pm$0.07 & $-0.24\pm$0.06 & $-2.23\pm$4.98 & $7.48\pm$3.01\\
J034030.72+242914.2 & TD035052N235741K01 & 55.128 & 24.48731 & 5249$\pm$20 & $4.76\pm$0.03 & $0.11\pm$0.01 & $-1.45\pm$1.6 & $5.76\pm$1.84\\
J034031.01+242141.0 & TD035052N235741K01 & 55.12922 & 24.3614 & 3789$\pm$8 & $4.67\pm$0.03 & $-0.65\pm$0.09 & $8.5\pm$1.59 & $15.61\pm$1.18\\
J034031.67+234521.9 & TD035052N235741K01 & 55.13198 & 23.75611 & 5868$\pm$65 & $4.29\pm$0.12 & $-0.12\pm$0.05 & $-22.59\pm$3.77 & $-11.75\pm$1.89\\
J034031.88+243419.1 & TD035052N235741K01 & 55.13285 & 24.57198 & 4876$\pm$96 & $3.03\pm$0.16 & $0.0\pm$0.06 & $46.06\pm$2.5 & $53.18\pm$2.59\\
J034034.36+234057.3 & TD035052N235741K01 & 55.1432 & 23.68261 & 5708$\pm$27 & $4.56\pm$0.02 & $0.21\pm$0.02 & $-6.38\pm$1.94 & $3.53\pm$1.66\\
J034034.76+232540.2 & TD035052N235741K01 & 55.14486 & 23.42786 & 7897$\pm$225 & $4.04\pm$0.16 & $-0.0\pm$0.13 & $-42.34\pm$8.23 & $-33.05\pm$7.58\\
J034034.92+243247.2 & TD035052N235741K01 & 55.14552 & 24.54648 & 6451$\pm$250 & $4.41\pm$0.21 & $0.32\pm$0.03 & $-31.89\pm$8.02 & $-25.27\pm$7.67\\
  \noalign{\smallskip}\hline
\end{tabular}
\smallskip\\
{This table is available in its entirety in machine-readable and Virtual Observatory (VO) forms in the online journal. A portion is shown here for guidance regarding its form and content.}
\end{table*}

\begin{table*}
\caption[]{Stellar parameters and RV from LASP estimation with MRS data. \label{par_houmed.tab}}
\setlength{\tabcolsep}{2pt}
\footnotesize
 \begin{tabular}{ccccccccc}
  \hline\noalign{\smallskip}
Name & Field & R.A. &  Dec. & $T_{\rm eff}$ &  log$g$  &  [Fe/H]  &  RV  &  corrected RV\\
 &  & (deg) &  (deg) & (K) &   &    &  (km/s)  &  (km/s) \\
   \hline\noalign{\smallskip}
J034004.12+235200.0 & TD035052N235741K01 & 55.0172 & 23.86668 & 6405$\pm$50 & $4.18\pm$0.03 & $-0.37\pm$0.04 & $-10.95\pm$0.35 & $-9.92\pm$1.4\\
J034007.72+241820.5 & TD035052N235741K01 & 55.03221 & 24.3057 & 5908$\pm$162 & $4.08\pm$0.23 & $-0.31\pm$0.08 & $5.79\pm$0.46 & $6.24\pm$0.52\\
J034008.18+241703.1 & TD035052N235741K01 & 55.03411 & 24.2842 & 4549$\pm$20 & $2.54\pm$0.05 & $0.09\pm$0.02 & $-39.6\pm$0.25 & $-39.2\pm$0.22\\
J034012.25+234313.8 & TD035052N235741K01 & 55.05106 & 23.72051 & 4894$\pm$28 & $2.31\pm$0.1 & $-0.5\pm$0.04 & $-25.81\pm$0.57 & $-24.78\pm$0.75\\
J034012.43+233803.1 & TD035052N235741K01 & 55.05184 & 23.63421 & 6353$\pm$61 & $4.14\pm$0.05 & $-0.27\pm$0.03 & $7.2\pm$2.31 & $8.72\pm$2.14\\
J034020.87+234005.1 & TD035052N235741K01 & 55.087 & 23.66809 & 5879$\pm$16 & $4.2\pm$0.02 & $-0.02\pm$0.01 & $40.25\pm$12.9 & $41.3\pm$12.15\\
J034024.32+242932.1 & TD035052N235741K01 & 55.10136 & 24.49227 & 4887$\pm$79 & $3.17\pm$0.08 & $-0.52\pm$0.04 & $-95.57\pm$2.04 & $-94.88\pm$2.04\\
J034025.96+232013.5 & TD035052N235741K01 & 55.10817 & 23.33711 & 6014$\pm$87 & $4.13\pm$0.11 & $-0.42\pm$0.06 & $60.65\pm$0.79 & $61.46\pm$1.29\\
J034029.22+234840.1 & TD035052N235741K01 & 55.1218 & 23.81117 & 5587$\pm$14 & $4.65\pm$0.02 & $0.01\pm$0.01 & $60.77\pm$0.3 & $61.74\pm$0.71\\
J034030.72+242914.2 & TD035052N235741K01 & 55.128 & 24.48731 & 5202$\pm$52 & $4.72\pm$0.06 & $0.06\pm$0.03 & $3.66\pm$0.41 & $4.04\pm$0.38\\
J034031.01+242141.0 & TD035052N235741K01 & 55.12922 & 24.3614 & 3769$\pm$9 & $4.64\pm$0.04 & $-0.82\pm$0.06 & $14.45\pm$0.22 & $14.91\pm$0.25\\
J034031.67+234521.9 & TD035052N235741K01 & 55.13198 & 23.75611 & 5894$\pm$126 & $4.34\pm$0.14 & $-0.18\pm$0.09 & $-14.2\pm$1.09 & $-13.64\pm$1.13\\
J034031.88+243419.1 & TD035052N235741K01 & 55.13285 & 24.57198 & 4768$\pm$25 & $3.06\pm$0.08 & $-0.11\pm$0.02 & $53.81\pm$0.35 & $54.19\pm$0.43\\
J034034.36+234057.3 & TD035052N235741K01 & 55.1432 & 23.68261 & 5673$\pm$10 & $4.54\pm$0.02 & $0.16\pm$0.01 & $4.73\pm$0.29 & $5.62\pm$1.09\\
J034035.40+232248.3 & TD035052N235741K01 & 55.14757 & 23.38009 & 5580$\pm$55 & $3.79\pm$0.11 & $0.24\pm$0.06 & $27.69\pm$1.2 & $28.98\pm$0.5\\
J034038.79+242507.8 & TD035052N235741K01 & 55.16163 & 24.41884 & 6345$\pm$52 & $4.07\pm$0.06 & $-0.09\pm$0.04 & $5.3\pm$0.36 & $5.77\pm$0.45\\
J034039.96+235046.7 & TD035052N235741K01 & 55.16651 & 23.84634 & 4801$\pm$36 & $2.65\pm$0.08 & $-0.25\pm$0.05 & $32.83\pm$1.16 & $34.08\pm$0.26\\
J034041.11+235922.0 & TD035052N235741K01 & 55.1713 & 23.98947 & 4572$\pm$46 & $1.88\pm$0.12 & $-0.69\pm$0.06 & $29.06\pm$1.94 & $30.65\pm$1.13\\
J034044.91+243926.0 & TD035052N235741K01 & 55.18717 & 24.65728 & 6845$\pm$32 & $4.17\pm$0.09 & $-0.28\pm$0.06 & $-21.52\pm$0.46 & $-21.08\pm$0.49\\
J034046.76+241255.7 & TD035052N235741K01 & 55.19484 & 24.2155 & 6261$\pm$28 & $4.3\pm$0.03 & $-0.05\pm$0.02 & $27.23\pm$0.36 & $27.65\pm$0.34\\
  \noalign{\smallskip}\hline
\end{tabular}
\smallskip\\
{This table is available in its entirety in machine-readable and Virtual Observatory (VO) forms in the online journal. A portion is shown here for guidance regarding its form and content.}
\end{table*}

\begin{table*}
\caption[]{Stellar parameters and RV from DD-Payne estimation with LRS data. \label{par_xianglow.tab}}
\setlength{\tabcolsep}{2pt}
\footnotesize
 \begin{tabular}{ccccccccc}
  \hline\noalign{\smallskip}
Name & Field & R.A. &  Dec. & $T_{\rm eff}$ &  log$g$  &  [Fe/H]  &  RV$_{\rm b}$  &  corrected RV\\
 &  & (deg) &  (deg) & (K) &   &    &  (km/s)  &  (km/s) \\
   \hline\noalign{\smallskip}
J034004.12+235200.0 & TD035052N235741K01 & 55.0172 & 23.86668 & 6351$\pm$16 & $4.14\pm$0.06 & $-0.44\pm$0.03 & $-29.17\pm$7.35 & $-17.0\pm$6.24\\
J034007.72+241820.5 & TD035052N235741K01 & 55.03221 & 24.3057 & 5772$\pm$52 & $3.75\pm$0.19 & $-0.41\pm$0.19 & $-0.59\pm$5.17 & $4.04\pm$5.02\\
J034008.18+241703.1 & TD035052N235741K01 & 55.03411 & 24.2842 & 4535$\pm$16 & $2.26\pm$0.08 & $0.07\pm$0.02 & $-44.2\pm$2.6 & $-40.1\pm$2.04\\
J034012.25+234313.8 & TD035052N235741K01 & 55.05106 & 23.72051 & 4911$\pm$27 & $2.47\pm$0.06 & $-0.46\pm$0.02 & $-35.17\pm$3.31 & $-23.52\pm$2.38\\
J034012.43+233803.1 & TD035052N235741K01 & 55.05184 & 23.63421 & 6177$\pm$12 & $3.95\pm$0.05 & $-0.38\pm$0.03 & $1.7\pm$7.96 & $13.24\pm$6.32\\
J034020.87+234005.1 & TD035052N235741K01 & 55.087 & 23.66809 & 5880$\pm$6 & $4.23\pm$0.02 & $-0.02\pm$0.01 & $40.0\pm$8.5 & $51.95\pm$10.17\\
J034020.90+242455.5 & TD035052N235741K01 & 55.08714 & 24.41547 & 5912$\pm$40 & $4.48\pm$0.19 & $-0.66\pm$0.06 & $-8.68\pm$5.79 & $-5.34\pm$5.83\\
J034024.32+242932.1 & TD035052N235741K01 & 55.10136 & 24.49227 & 4896$\pm$24 & $2.91\pm$0.05 & $-0.52\pm$0.03 & $-98.56\pm$4.27 & $-94.46\pm$3.58\\
J034025.64+244209.5 & TD035052N235741K01 & 55.10687 & 24.70268 & 6278$\pm$58 & $3.74\pm$0.13 & $-0.39\pm$0.05 & $-14.98\pm$23.51 & $-11.41\pm$24.06\\
J034025.96+232013.5 & TD035052N235741K01 & 55.10817 & 23.33711 & 5907$\pm$31 & $4.0\pm$0.05 & $-0.52\pm$0.03 & $51.55\pm$6.72 & $63.68\pm$6.86\\
J034026.48+235823.3 & TD035052N235741K01 & 55.11036 & 23.97316 & 4583$\pm$49 & $2.27\pm$0.09 & $-0.54\pm$0.55 & $-51.4\pm$24.51 & $-40.39\pm$24.31\\
J034029.22+234840.1 & TD035052N235741K01 & 55.1218 & 23.81117 & 5491$\pm$10 & $4.46\pm$0.03 & $-0.08\pm$0.02 & $49.64\pm$2.82 & $61.99\pm$1.31\\
J034029.59+233303.9 & TD035052N235741K01 & 55.12332 & 23.55111 & 4396$\pm$41 & $4.5\pm$0.08 & $-0.43\pm$0.07 & $-6.31\pm$9.38 & $4.89\pm$9.01\\
J034030.72+242914.2 & TD035052N235741K01 & 55.128 & 24.48731 & 5137$\pm$7 & $4.53\pm$0.04 & $-0.07\pm$0.02 & $0.1\pm$2.26 & $5.08\pm$2.97\\
J034031.01+242141.0 & TD035052N235741K01 & 55.12922 & 24.3614 & 4260$\pm$26 & $4.33\pm$0.1 & $-0.92\pm$0.05 & $0.58\pm$8.8 & $4.7\pm$8.27\\
J034031.67+234521.9 & TD035052N235741K01 & 55.13198 & 23.75611 & 5836$\pm$32 & $4.23\pm$0.12 & $-0.23\pm$0.06 & $-25.68\pm$9.39 & $-13.17\pm$8.51\\
J034031.88+243419.1 & TD035052N235741K01 & 55.13285 & 24.57198 & 4791$\pm$48 & $2.76\pm$0.07 & $-0.06\pm$0.04 & $49.55\pm$6.33 & $53.71\pm$6.57\\
J034034.36+234057.3 & TD035052N235741K01 & 55.1432 & 23.68261 & 5586$\pm$8 & $4.46\pm$0.02 & $0.04\pm$0.02 & $-8.51\pm$2.12 & $2.99\pm$1.84\\
J034034.76+232540.2 & TD035052N235741K01 & 55.14486 & 23.42786 & 7667$\pm$85 & $4.06\pm$0.24 & $-0.3\pm$0.13 & $-45.06\pm$9.66 & $-34.17\pm$9.36\\
J034034.92+243247.2 & TD035052N235741K01 & 55.14552 & 24.54648 & 6446$\pm$190 & $4.34\pm$0.29 & $0.08\pm$0.12 & $13.76\pm$40.23 & $17.07\pm$39.8\\
  \noalign{\smallskip}\hline
\end{tabular}
\smallskip\\
{This table is available in its entirety in machine-readable and Virtual Observatory (VO) forms in the online journal. A portion is shown here for guidance regarding its form and content.}
\end{table*}

\begin{table*}
\caption[]{Stellar parameters and RV from SLAM estimation with MRS data. \label{par_zhangmed.tab}}
\setlength{\tabcolsep}{2pt}
\footnotesize
 \begin{tabular}{ccccccccc}
  \hline\noalign{\smallskip}
Name & Field & R.A. &  Dec. & $T_{\rm eff}$ &  log$g$  &  [Fe/H]  &  RV  &  corrected RV\\
 &  & (deg) &  (deg) & (K) &   &    &  (km/s)  &  (km/s) \\
   \hline\noalign{\smallskip}
J034004.12+235200.0 & TD035052N235741K01 & 55.0172 & 23.86668 & 6323$\pm$103 & $4.16\pm$0.14 & $-0.48\pm$0.05 & $-10.39\pm$0.28 & $-10.07\pm$0.99\\
J034007.72+241820.5 & TD035052N235741K01 & 55.03221 & 24.3057 & 5811$\pm$156 & $4.11\pm$0.18 & $-0.23\pm$0.09 & $6.41\pm$0.42 & $6.22\pm$0.41\\
J034008.18+241703.1 & TD035052N235741K01 & 55.03411 & 24.2842 & 4551$\pm$70 & $1.99\pm$0.11 & $0.11\pm$0.05 & $-38.57\pm$0.23 & $-38.78\pm$0.18\\
J034012.25+234313.8 & TD035052N235741K01 & 55.05106 & 23.72051 & 4924$\pm$74 & $2.16\pm$0.15 & $-0.51\pm$0.06 & $-25.57\pm$0.3 & $-25.31\pm$0.71\\
J034012.43+233803.1 & TD035052N235741K01 & 55.05184 & 23.63421 & 6092$\pm$89 & $3.93\pm$0.09 & $-0.38\pm$0.06 & $7.82\pm$2.31 & $8.57\pm$2.16\\
J034020.87+234005.1 & TD035052N235741K01 & 55.087 & 23.66809 & 5874$\pm$37 & $4.15\pm$0.04 & $-0.05\pm$0.03 & $40.95\pm$12.68 & $41.27\pm$12.07\\
J034024.32+242932.1 & TD035052N235741K01 & 55.10136 & 24.49227 & 4812$\pm$162 & $2.73\pm$0.49 & $-0.65\pm$0.15 & $-94.08\pm$0.5 & $-94.15\pm$0.42\\
J034025.96+232013.5 & TD035052N235741K01 & 55.10817 & 23.33711 & 5865$\pm$90 & $4.15\pm$0.12 & $-0.47\pm$0.08 & $61.19\pm$0.69 & $61.29\pm$1.12\\
J034029.22+234840.1 & TD035052N235741K01 & 55.1218 & 23.81117 & 5444$\pm$46 & $4.55\pm$0.05 & $-0.18\pm$0.03 & $61.44\pm$0.29 & $61.7\pm$0.61\\
J034030.72+242914.2 & TD035052N235741K01 & 55.128 & 24.48731 & 5076$\pm$75 & $4.38\pm$0.12 & $-0.14\pm$0.03 & $4.39\pm$0.42 & $4.16\pm$0.4\\
J034031.01+242141.0 & TD035052N235741K01 & 55.12922 & 24.3614 & 3337$\pm$98 & $2.72\pm$0.23 & $-0.89\pm$0.12 & $15.99\pm$0.3 & $15.8\pm$0.31\\
J034031.67+234521.9 & TD035052N235741K01 & 55.13198 & 23.75611 & 5882$\pm$39 & $4.28\pm$0.15 & $-0.37\pm$0.11 & $-13.31\pm$0.68 & $-13.55\pm$0.68\\
J034031.88+243419.1 & TD035052N235741K01 & 55.13285 & 24.57198 & 4691$\pm$136 & $2.65\pm$0.23 & $-0.14\pm$0.05 & $54.29\pm$0.41 & $54.09\pm$0.44\\
J034034.36+234057.3 & TD035052N235741K01 & 55.1432 & 23.68261 & 5471$\pm$40 & $4.4\pm$0.05 & $0.06\pm$0.03 & $5.43\pm$0.34 & $5.62\pm$0.93\\
J034035.40+232248.3 & TD035052N235741K01 & 55.14757 & 23.38009 & 5510$\pm$96 & $3.7\pm$0.21 & $0.2\pm$0.1 & $28.7\pm$1.21 & $29.18\pm$0.49\\
J034038.79+242507.8 & TD035052N235741K01 & 55.16163 & 24.41884 & 6200$\pm$75 & $3.83\pm$0.08 & $-0.16\pm$0.07 & $5.98\pm$0.3 & $5.82\pm$0.34\\
J034039.96+235046.7 & TD035052N235741K01 & 55.16651 & 23.84634 & 4835$\pm$76 & $2.4\pm$0.12 & $-0.26\pm$0.06 & $33.93\pm$0.95 & $34.41\pm$0.23\\
J034041.11+235922.0 & TD035052N235741K01 & 55.1713 & 23.98947 & 4719$\pm$81 & $1.31\pm$0.14 & $-0.79\pm$0.08 & $29.64\pm$0.79 & $30.46\pm$0.16\\
J034044.91+243926.0 & TD035052N235741K01 & 55.18717 & 24.65728 & 6309$\pm$254 & $3.9\pm$0.35 & $-0.48\pm$0.1 & $-21.46\pm$0.72 & $-21.63\pm$0.69\\
J034046.76+241255.7 & TD035052N235741K01 & 55.19484 & 24.2155 & 6128$\pm$46 & $4.19\pm$0.05 & $-0.09\pm$0.03 & $27.72\pm$0.3 & $27.51\pm$0.26\\
  \noalign{\smallskip}\hline
\end{tabular}
\smallskip\\
{This table is available in its entirety in machine-readable and Virtual Observatory (VO) forms in the online journal. A portion is shown here for guidance regarding its form and content.}
\end{table*}

\end{document}